\documentclass[10pt, journal]{IEEEtran}

\usepackage{graphicx, bm, mathrsfs}
\usepackage{cite,array}
\usepackage{amssymb, dsfont, color}
\usepackage{epstopdf}
\usepackage[cmex10]{amsmath}
\usepackage{enumerate}
\usepackage{multirow}
\usepackage{booktabs}
\usepackage{threeparttable}
\usepackage{subeqnarray}
\usepackage{cases, stmaryrd}
\usepackage{stackengine}
\newcommand\xrowht[2][0]{\addstackgap[.5\dimexpr#2\relax]{\vphantom{#1}}}

\newtheorem{assumption}{Assumption}
\newtheorem{theorem}{Theorem}

\newtheorem{proposition}{Proposition}
\newtheorem{definition}{Definition}
\newtheorem{remark}{Remark}

\newcommand{\p}{\textrm{p}}
\newcommand{\rf}{\textrm{r}}
\newcommand{\ct}{\textrm{c}}
\newcommand{\f}{\textrm{f}}
\newcommand{\ob}{\textrm{o}}
\newcommand{\ax}{\textrm{a}}
\newcommand{\dom}{\textrm{dom}}
\DeclareMathOperator*{\esssup}{\textrm{ess.\,sup}}

\newcommand{\diag}{\textrm{diag}}

\newcommand{\tran}{\mathrm{T}}
\newcommand{\overbar}[1]{\mkern 1.5mu\overline{\mkern-1.5mu#1\mkern-1.5mu}\mkern 1.5mu}

\begin{document}

\title{\LARGE \bf Event-Triggered Tracking Control of Networked Multi-Agent Systems
\thanks{This work was supported by the H2020 ERC Consolidator Grants L2C (864017) and LEAFHOUND (864720), the CHIST-ERA 2018 project DRUID-NET, the Swedish Research Council (VR) and the Knut och Alice Wallenberg Foundation (KAW).}
}

\author{Wei~Ren, and Dimos V. Dimarogonas, \IEEEmembership{Senior Member, IEEE}
\thanks{W. Ren is with ICTEAM institute, Universit\'{e} catholique de Louvain, 1348 Louvain-la-Neuve, Belgium. D. V. Dimarogonas is with Division of Decision and Control Systems, EECS, KTH Royal Institute of Technology, SE-10044, Stockholm, Sweden.
Email: \texttt{\small w.ren@uclouvain.be}, \texttt{\small dimos@kth.se}.}
}

\maketitle

\begin{abstract}
This paper studies the tracking control problem of networked multi-agent systems under both multiple networks and event-triggered mechanisms. Multiple networks are to connect multiple agents and reference systems with decentralized controllers to guarantee their information transmission, whereas the event-triggered mechanisms are to reduce the information transmission via the networks. In this paper, each agent has a network to communicate with its controller and reference system, and all networks are independent and asynchronous and have local event-triggered mechanisms, which are based on local measurements and determine whether the local measurements need to be transmitted via the corresponding network. To address this scenario, we first implement the emulation-based approach to develop a novel hybrid model for the tracking control of networked multi-agent systems. Next, sufficient conditions are derived and decentralized event-triggered mechanisms are designed to guarantee the desired tracking performance. Furthermore, the proposed approach is applied to derive novel results for the event-triggered observer design problem of networked multi-agent systems. Finally, two numerical examples are presented to illustrate the validity of the developed results.
\end{abstract}


\section{Introduction}
\label{sec-introduction}

In the era of the Internet of Things, smart devices are able to interconnect and interplay to link the physical world to the digital world \cite{Al2015internet}. The sensing, communication, computation and control are integrated into different levels of operations and information. In particular, the introduction of wired/wireless networks to connect multiple agents leads to networked multi-agent systems (MAS). The presence of networks improves efficiency and flexibility of integrated applications, and reduces installation and maintenance time and costs \cite{Gupta2010networked, Heemels2010networked, Dolk2016output}. Since multiple agents are physically distributed and interconnected to coordinate their tasks and to achieve overall specifications, cooperative control of MAS has attracted numerous attention from various communities \cite{Olfati2007consensus, Ren2008distributed}. The main challenge in cooperative control is how to design control schemes to limit transmission delays and packet dropouts to avoid the deterioration of the desired performances and to achieve an agreement for multiple agents by exploiting information from each agent and its neighbors. One attractive approach in this context is periodic event-triggered control (PETC) \cite{Heemels2012periodic, Garcia2016periodic, Yu2018explicit}, combining time-triggered control (TTC) (where the information is transmitted at discrete-time instants \cite{Poveda2019hybrid, Gao2011sampled}) and event-triggered control (ETC) (where the information is transmitted only when the triggering condition is satisfied \cite{Tabuada2007event, Mazo2011decentralized, Dimarogonas2011distributed}). In the PETC, the triggering condition is evaluated with a predefined sampling period to decide the information transition, thereby resulting in a balance between TTC and ETC to avoid the continuous evaluation of the triggering condition \cite{Heemels2012periodic, Yu2018explicit}.

Many existing results on cooperative control of (networked) MAS focus mainly on consensus to a common point. Both TTC and ETC/PETC have been addressed \cite{Mazo2011decentralized, Dimarogonas2011distributed, Seyboth2013event}. However, tracking control, as a fundamental problem in control theory \cite{Ren2019tracking, Postoyan2014tracking}, received few attention \cite{Hong2006tracking, Cheng2016event, Postoyan2015event}. The main objective of the tracking control is to design controllers such that multiple agents can track the given reference trajectories as close as possible \cite{Tallapragada2013event, Biemond2013tracking}. In the tracking control, the controller consists of two parts \cite{Van2010tracking}: the feedforward part to induce the reference trajectories for the agents, and the feedback part to drive the agents to converge to the reference trajectories. In MAS, each agent only has the local information from its neighbor agents, while being able to take actions independently without having to wait for a central control signal. These properties affect many system performances including the tracking performance, and thus may result in some challenges for the tracking control. Besides, as opposed to the traditional tracking problem, another main challenge of the tracking control of networked MAS is that only local/partial information is transmitted to the agents due to the limited capacity of communication networks. The information transmission via networks may be an error source affecting the tracking performance \cite{Van2010tracking}. From the above discussion, both network-induced errors and local interaction rules need to be considered simultaneously, which results in the main difficulties in tracking performance analysis.

In this paper, we study the event-triggered tracking control problem for networked MAS. To this end, we implement an emulation-like approach as in \cite{Ren2019tracking, Postoyan2014tracking, Heemels2010networked} and develop a novel hybrid model using the formalism in \cite{Cai2009characterizations, Goebel2012hybrid} to address the tracking control for networked MAS, which is our first contribution. Specifically, a general scenario is considered: multiple independent and asynchronous networks are applied to ensure the communication among different components. Such setting is reasonable due to the connection among sensors, controllers and actuators via different communication channels, and allows to recover the architectures in \cite{Ren2019tracking, Postoyan2014tracking, Van2010tracking} for networked control systems (NCS) and \cite{Cheng2016event, Postoyan2015event} for MAS as particular cases. Based on this setting, a general hybrid model is developed to incorporate all cases caused by multiple network and decentralized event-triggered mechanisms (ETMs), which further lead to different types of jumps in the developed hybrid model. To investigate these types of jumps and the network-induced errors, a novel Lyapunov function is proposed for the tracking performance analysis. Furthermore, both Lyapunov-based conditions and decentralized ETMs are derived. The tradeoff between the maximally allowable sampling period (MASP) and the maximally allowable delay (MAD) is derived to guarantee the convergence of the tracking errors with respect to the external disturbance and the network-induced errors.

Since the controller design is usually based on the state estimation \cite{Hong2008distributed, Park2016design, Huang2017toward} and the observer design can be connected with the tracking control in terms of synchronization \cite{Postoyan2014tracking, Nijmeijer1997observer}, our second contribution is to apply the derived results to the event-triggered observer design for networked MAS. To show this, we address the following two cases: an independent observer is designed for each agent based on the local information from this agent; multiple observers are designed for an agent (group) based on the partial information from the agent (group). In these two cases, the agents are not required to be stable (as in \cite{Postoyan2011framework}) due to their own nature, and the robust decentralized estimation is ensured under the derived decentralized ETMs and MASP bounds. In particular, we envision a hybrid model similar to \cite{Li2017robust}, which however is on the linear time-triggered case. Therefore, the obtained results are novel in the context of robust distributed estimation, in which case the estimation is based on either TTC/ETC or the centralized fashion \cite{Postoyan2014tracking, Hong2008distributed, Shoukry2015event, Huang2017toward}.

The remainder of this paper is organized below. Preliminaries are given in Section \ref{sec-preliminaries}. The tracking problem is formulated in Section \ref{sec-problemformation}, and the hybrid model is developed in Section \ref{sec-developandreformulate}. The main results are derived in Section \ref{sec-mainresults}. The obtained results are applied to the event-triggered observer design in Section \ref{sec-observer}. Numerical examples are presented in Section \ref{sec-illustration}. Conclusions and further research are stated in Section \ref{sec-conclusion}.

\section{Preliminaries}
\label{sec-preliminaries}

$\mathbb{R}:=(-\infty, +\infty)$; $\mathbb{R}_{\geq0}:=[0, +\infty)$; $\mathbb{R}_{>0}:=(0, +\infty)$; $\mathbb{N}:=\{0, 1, 2, \ldots\}$; $\mathbb{N}_{+}:=\{1, 2, \ldots\}$. Given two sets $\mathcal{A}$ and $\mathcal{B}$, $\mathcal{B}\backslash\mathcal{A}:=\{x: x\in\mathcal{B}, x\notin\mathcal{A}\}$. $|\cdot|$ denotes the Euclidean norm. Given two vectors $x, y\in\mathbb{R}^{n}$, $(x, y):=(x^{\top}, y^{\top})^{\top}$ for simplicity of notation, and $\langle x, y\rangle$ denotes the usual inner product. $\mathds{E}$ denotes the vector with all components being 1, $I$ denotes the identity matrix of appropriate dimension, and $\diag\{A, B\}$ denotes the block diagonal matrix made of the matrices $A$ and $B$. Given a function $f: \mathbb{R}_{\geq t_{0}}\rightarrow\mathbb{R}^{n}$, $f(t^{+}):=\limsup_{s\rightarrow0^{+}}f(t+s)$. A function $\alpha: \mathbb{R}_{\geq0}\rightarrow\mathbb{R}_{\geq0}$ is of class $\mathcal{K}$ if it is continuous, $\alpha(0)=0$, and strictly increasing; it is of class $\mathcal{K}_{\infty}$ if it is of class $\mathcal{K}$ and unbounded; it is of class $\mathcal{L}$ if it is continuous, strictly decreasing, and $\lim_{s\rightarrow\infty}\alpha(s)=0$. A function $\beta: \mathbb{R}^{2}_{\geq0}\rightarrow\mathbb{R}_{\geq0}$ is of class $\mathcal{KL}$ if $\beta(s, t)\in\mathcal{K}$ for fixed $t\geq0$ and $\beta(s, t)\in\mathcal{L}$ for fixed $s\geq0$. A function $\beta:\mathbb{R}^{3}_{\geq0}\rightarrow\mathbb{R}_{\geq0}$ is of class $\mathcal{KLL}$ if $\beta(r, s, t)\in\mathcal{KL}$ for fixed $s\geq0$ and $\beta(r, s, t)\in\mathcal{KL}$ for fixed $t\geq0$.

\subsection{Hybrid System}
\label{subsec-hybsys}

The basic concepts of hybrid systems are introduced below; see \cite{Cai2009characterizations} for the details. Consider the following hybrid system
\begin{align}
\label{eqn-1}
\left\{\begin{aligned}
&\dot{x}=F(x, w), &\quad& (x, w)\in C,  \\
&x^{+}=G(x, w), &\quad& (x, w)\in D,
\end{aligned}\right.
\end{align}
where $x\in\mathbb{R}^{n}$ is the system state, $w\in\mathbb{R}^{m}$ is the external input, $F: C\rightarrow\mathbb{R}^{n}$ is the flow map, $G: D\rightarrow\mathbb{R}^{m}$ is the jump map, $C$ is the flow set and $D$ is the jump set. For the hybrid system \eqref{eqn-1}, the following basic assumptions are presented: the sets $C, D\subset\mathbb{R}^{n}\times\mathbb{R}^{m}$ are closed; $F$ is continuous on $C$; and $G$ is continuous on $D$. In \eqref{eqn-1}, $x\in\mathbb{R}^{n}$ is defined on \emph{hybrid time domain}, which is denoted by $\dom x\subset\mathbb{R}_{\geq0}\times\mathbb{N}$ with the following structure: for each $(T, J)\in\dom x$, $\dom x\cap([0, T]\times\{0, \ldots, J\})$ can be written as $\bigcup_{0\leq j\leq J-1}([t_{j}, t_{j+1}], j)$ for some finite sequence of times $0=t_{0}\leq t_{1}\leq\ldots\leq t_{J}$. Given $(t_{1}, j_{1}), (t_{2}, j_{2})\in\mathbb{R}_{\geq0}\times\mathbb{N}$, we denote by $(t_{1}, j_{1})\preceq(t_{2}, j_{2})$ (or $(t_{1}, j_{1})\prec(t_{2}, j_{2})$) if $t_{1}+j_{1}\leq t_{2}+j_{2}$ (or $t_{1}+j_{1}<t_{2}+j_{2}$). A solution $(x, w)$ to \eqref{eqn-1} is a function on the hybrid time domain satisfying the dynamics in \eqref{eqn-1} with the following property: $\dom x=\dom w$; $x(\cdot, j)$ with fixed $j$ is absolutely continuous; and $w(\cdot, j)$ with fixed $j$ is Lebesgue measurable and locally essentially bounded. A solution $(x, w)$ is \emph{maximal} if it cannot be extended. Define $\|w\|_{(t, j)}:=\max\left\{\esssup\limits_{(t', j')\in\dom w\setminus\Xi(w), (0, 0)\preceq(t', j')\preceq(t, j)}|w(t', j')|,\right.$ $\left.\sup\limits_{(t, j)\in\Xi(w), (0, 0)\preceq(t', j')\preceq(t, j)}\sup|w(t', j')|\right\}$ where $\Xi(w):=$ $\{(t, j)\in\dom w: (t, j+1)\in\dom w\}$. $\mathfrak{S}_{w}(x_{0})$ denotes the set of all maximal solutions to \eqref{eqn-1} with $x_{0}=x(0, 0)\in C\cup D$ and finite $\|w\|:=\sup_{(t, j)\in\dom w}\|w\|_{(t, j)}$.

\begin{definition}[\cite{Cai2009characterizations}]
The hybrid system \eqref{eqn-1} is \emph{input-to-state stable (ISS) } from $w$ to $x$, if there exist $\beta\in\mathcal{KLL}$ and $\gamma\in\mathcal{K}_{\infty}$ such that $|x(t, j)|\leq\beta(|x(0, 0)|,t, j)+\gamma(\|w\|_{(t, j)})$ for all $(t, j)\in\dom x$ and all $(x, w)\in\mathfrak{S}_{w}(x_{0})$.
\end{definition}

\subsection{Graph Theory}
\label{subsec-graph}

A directed graph is defined as $\mathcal{G}=\{\mathcal{V}, \mathcal{E}\}$, where $\mathcal{V}=\{1, 2, \ldots, N\}$ is the set of nodes and $\mathcal{E}\subseteq\mathcal{V}\times\mathcal{V}$ the set of edges. Each edge directly links two nodes. An edge from node $i$ to node $j$ is denoted by $(i, j)$, and implies that node $i$ can receive information from node $j$. The \emph{adjacency matrix} is denoted by $\mathcal{A}=[a_{ij}]_{N\times N}$, where $a_{ij}=1$ if $(i, j)\in\mathcal{E}$ and $a_{ij}=0$ otherwise. The neighbor set of node $i$ is denoted by $\mathcal{N}_{i}=\{j\in\mathcal{V}: (i, j)\in\mathcal{E}\}$. The directed graph $\mathcal{G}$ is \emph{undirected} if $a_{ij}=a_{ji}$ for all $i, j\in\mathcal{V}$; and $\mathcal{G}$ is \emph{connected} if for all $i, j\in\mathcal{V}$, there exists a path connecting $i$ and $j$, which is an ordered list of edges, i.e., $(i, k_{1})(k_{1}, k_{2})\ldots(k_{2}, k_{\mathfrak{n}})(k_{\mathfrak{n}}, j)$ with finite $\mathfrak{n}\in\mathbb{N}_{+}$. A directed graph is \emph{all-to-all connected} if two distinct nodes are connected by a unique edge.

\section{Problem Formulation}
\label{sec-problemformation}

In this section, we first state the tracking control problem for networked MAS studied in this paper, and then present the detailed information transmission among all agents, references and controllers via multiple networks.

\subsection{Tracking Problem of Networked MAS}
\label{subsec-trackingproblem}

Consider the nonlinear MAS with $N\in\mathbb{N}_{+}$ agents, whose dynamics is described as
\begin{align}
\label{eqn-2}
\dot{x}^{i}_{\p}=f^{i}_{\p}(x_{\p}, u_{i}, w_{\p}),\quad y^{i}_{\p}=g^{i}_{\p}(x^{i}_{\p}),
\end{align}
where $i\in\mathcal{N}:=\{1, \ldots, N\}$; for the $i$-th agent, $x^{i}_{\p}\in\mathbb{R}^{n^{i}_{\p}}$ is the agent state, $u_{i}\in\mathbb{R}^{n^{i}_{u}}$ is the control input, $w_{\p}\in\mathbb{R}^{n_{1}}$ is the external disturbance, and $y^{i}_{\p}\in\mathbb{R}^{n^{i}_{y}}$ is the output. For the $i$-th agent, its reference system to be tracked is given by
\begin{align}
\label{eqn-3}
\dot{x}^{i}_{\rf}=f^{i}_{\p}(x_{\rf}, u^{i}_{\f}, w_{\rf}),\quad y^{i}_{\rf}=g^{i}_{\p}(x^{i}_{\rf}),
\end{align}
where $x^{i}_{\rf}\in\mathbb{R}^{n^{i}_{\p}}$ is the reference state, $u^{i}_{\f}\in\mathbb{R}^{n^{i}_{u}}$ is the feedforward control input of the reference system, $w_{\rf}\in\mathbb{R}^{n_{2}}$ is the external disturbance, and $y^{i}_{\rf}\in\mathbb{R}^{n^{i}_{y}}$ is the reference output.

\begin{figure}[!t]
\begin{center}
\begin{picture}(65, 90)
\put(-65, -12){\resizebox{65mm}{35mm}{\includegraphics[width=2.5in]{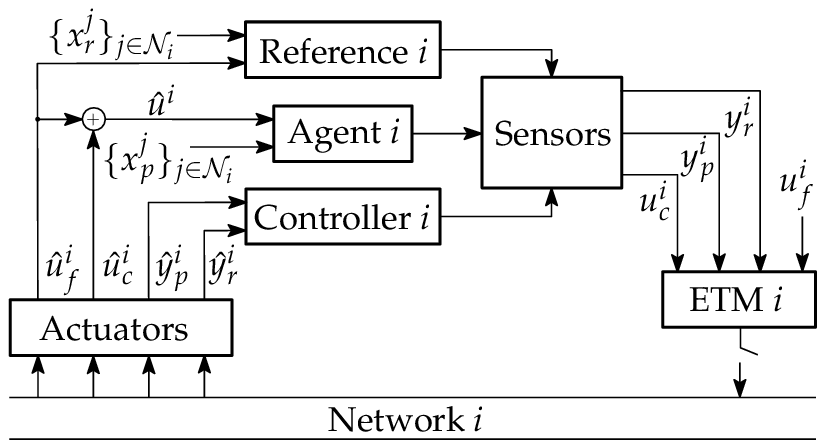}}}
\end{picture}
\end{center}
\caption{Illustration of the information transmission in a single network. $\{x^{j}_{\p}\}_{j\in\mathcal{N}_{i}}$ is the set of states from the $i$-th agent's neighbors, and $\{x^{j}_{\rf}\}_{j\in\mathcal{N}_{i}}$ is the set of states from the $i$-th reference's neighbors.}
\label{fig-1}
\end{figure}

Let $x_{\p}:=(x^{1}_{\p}, \ldots, x^{N}_{\p})\in\mathbb{R}^{n_{\p}}$ and  $x_{\rf}:=(x^{1}_{\rf}, \ldots, x^{N}_{\rf})\in\mathbb{R}^{n_{\p}}$ with $n_{\p}:=\sum^{N}_{i=1}n^{i}_{\p}$. In \eqref{eqn-2}-\eqref{eqn-3}, $f^{i}_{\p}$ (or $f^{i}_{\rf}$) is written as a function related to the overall vector $x_{\p}$ (or $x_{\rf}$), but it depends only on the states of the $i$-th agent (or reference) and its neighbor agents (or references); see Fig. \ref{fig-1}. The physical coupling among all agents is characterized by a graph $\mathcal{G}_{\p}:=(\mathcal{N}, \mathcal{E}_{\p})$, whereas the physical coupling among all references is characterized by a graph $\mathcal{G}_{\rf}:=(\mathcal{N}, \mathcal{E}_{\rf})$. For these two graphs, the following assumption is made; see also \cite{Poveda2019hybrid, Garcia2016periodic, Yu2018explicit}.

\begin{assumption}
\label{asn-1}
$\mathcal{G}_{\p}$ and $\mathcal{G}_{\rf}$ are undirected and connected.
\end{assumption}

The graphs $\mathcal{G}_{\p}$ and $\mathcal{G}_{\rf}$ are not necessarily the same, and will not play a central role in our results. To track the reference system \eqref{eqn-3}, assume that in the absence of the network, the controller for the $i$-th agent is designed as
\begin{equation}
\label{eqn-4}
u_{i}=u^{i}_{\ct}+u^{i}_{\f},
\end{equation}
where $u^{i}_{\ct}\in\mathbb{R}^{n^{i}_{u}}$ is the feedback control input, and is generated by the following feedback controller
\begin{align}
\label{eqn-5}
\dot{x}^{i}_{\ct}=f^{i}_{\ct}(x^{i}_{\ct}, y^{i}_{\p}, y^{i}_{\rf}, w_{\ct}),\quad u^{i}_{\ct}=g^{i}_{\ct}(x^{i}_{\ct}),
\end{align}
where $x^{i}_{\ct}\in\mathbb{R}^{n^{i}_{\ct}}$ is the controller state, and $w_{\ct}\in\mathbb{R}^{n_{3}}$ is the external disturbance. Therefore, in the absence of the network and under the controller \eqref{eqn-4}, $x^{i}_{\p}$ should converge to $x^{i}_{\rf}$ as close as possible. That is, the tracking goal is achieved, if
\begin{equation}
\label{eqn-6}
|x^{i}_{\p}(t)-x^{i}_{\rf}(t)|\leq\beta(|x^{i}_{\p}(0)-x^{i}_{\rf}(0)|, t)+\gamma(\|w\|),
\end{equation}
where $\beta\in\mathcal{KL}, \gamma\in\mathcal{K}_{\infty}$, and $w:=(w_{\p}, w_{\rf}, w_{\ct})\in\mathbb{R}^{n_{w}}$ with $n_{w}:=n_{1}+n_{2}+n_{3}$.

\begin{remark}
\label{rmk-1}
For each agent, the dynamics \eqref{eqn-2} is general and recovers the single/double integrators \cite{Mazo2011decentralized, Dimarogonas2011distributed, Seyboth2013event} and heterogenous models \cite{Cheng2016event, Kim2010output} as special cases. The controller \eqref{eqn-5} depends only on the corresponding agent and reference system, which implies that the controllers are not affected by other agents and reference systems. Such controller exists for both MAS \cite{Kim2010output, Wang2010distributed} and NCS \cite{Dolk2016output,  Freirich2016decentralized, Pola2017decentralized}.
\hfill $\square$
\end{remark}

We denote $x_{\ct}:=(x^{1}_{\ct}, \ldots, x^{N}_{\ct})\in\mathbb{R}^{n_{\ct}}$, $u:=(u_{1}, \ldots, u_{N})\in\mathbb{R}^{n_{u}}$, $u_{\f}:=(u^{1}_{\f}, \ldots, u^{N}_{\f})\in\mathbb{R}^{n_{u}}$, $u_{\ct}:=(u^{1}_{\ct}, \ldots, u^{N}_{\ct})\in\mathbb{R}^{n_{u}}$, $y_{\p}:=(y^{1}_{\p}, \ldots, y^{N}_{\p})\in\mathbb{R}^{n_{y}}$, and $y_{\rf}:=(y^{1}_{\rf}, \ldots, y^{N}_{\rf})\in\mathbb{R}^{n_{y}}$, where $n_{\ct}:=\sum^{N}_{i=1}n^{i}_{\ct}$, $n_{u}:=\sum^{N}_{i=1}n^{i}_{u}$ and $n_{y}:=\sum^{N}_{i=1}n^{i}_{y}$. Hence, the dynamics of all agents can be written unifiedly as
\begin{align}
\label{eqn-7}
\dot{x}_{\p}=f_{\p}(x_{\p}, u, w_{\p}),\quad y_{\p}=g_{\p}(x_{\p}),
\end{align}
where $f_{\p}:=(f^{1}_{\p}, \ldots, f^{N}_{\p})\in\mathbb{R}^{n_{\p}}$ and $g_{\p}:=(g^{1}_{\p}, \ldots, g^{N}_{\p})\in\mathbb{R}^{n_{y}}$. Accordingly, the reference system is given by
\begin{align}
\label{eqn-8}
\dot{x}_{\rf}=f_{\p}(x_{\rf}, u_{\f}, w_{\rf}),\quad y_{\rf}=g_{\p}(x_{\rf}).
\end{align}
Assume that the system \eqref{eqn-8} has a unique solution for any initial condition and any input. All controllers \eqref{eqn-4} are stacked as
\begin{equation}
\label{eqn-9}
u=u_{\ct}+u_{\f},
\end{equation}
and the feedback control input $u_{\ct}\in\mathbb{R}^{n_{u}}$ comes from the following feedback controller
\begin{align}
\label{eqn-10}
\dot{x}_{\ct}=f_{\ct}(x_{\ct}, y_{\p}, y_{\rf}, w_{\ct}),\quad u_{\ct}=g_{\ct}(x_{\ct}).
\end{align}
We assume that $f_{\p}$ and $f_{\ct}$ are continuous; $g_{\p}$ and $g_{\ct}$ are continuously differentiable.

Our objective is to implement the designed controller \eqref{eqn-9} over both ETMs and multiple networks, and to demonstrate that the assumed tracking performance of the system \eqref{eqn-7}-\eqref{eqn-10} will be preserved \emph{approximately} for the networked MAS under reasonable assumptions and the designed ETMs. To be specific, in the network-free case, the tracking performance in \eqref{eqn-6} for MAS is achieved under the controller \eqref{eqn-9}. However, in the networked case, \eqref{eqn-6} may be not achieved due to network-induced constraints, which will be introduced in the next subsection. In addition, to mitigate the unnecessary waste of communication resources, the ETM is designed for each network to balance resource utilization and control performance. Therefore, our goal is to establish conditions on both the networks and the system \eqref{eqn-7}-\eqref{eqn-10} and to design decentralized ETMs to guarantee the \emph{approximate} convergence of $x_{\p}$ towards $x_{\rf}$ in the presence of the network-induced constraints and designed ETMs. Here, `\emph{approximate}' means that the convergence region depends not only on the external disturbance as in \eqref{eqn-6} but also on the network-induced errors.

\subsection{Information Transmission over Multiple Networks}
\label{subsec-informationtransmission}

As shown in Fig. \ref{fig-1}, the information is sampled via the sensors and then determined (by the ETM to be designed in Section \ref{sec-mainresults}) to be transmitted via the network. For different agents and references, the information may be transmitted via different networks (e.g., wired/wireless networks \cite{Gupta2010networked, Dolk2016output}), and is transmitted only when the information is needed. Whether the information is needed is evaluated by the ETM. Hence, the information transmissions via multiple networks are independent and may not be synchronous.

\begin{assumption}
\label{asn-2}
In the case that the ETM is implemented, all sensors and actuators are connected via $N\in\mathbb{N}_{+}$ independent and asynchronous networks.
\end{assumption}

If some agents can be composed as a agent group sharing a common network, then the number of multiple networks can be reduced and be smaller than the number of the agents. The information to be transmitted is denoted by $z_{i}:=(y^{i}_{\p}, y^{i}_{\rf}, u^{i}_{\f}, u^{i}_{\ct})\in\mathbb{R}^{n^{i}_{z}}$ with $n^{i}_{z}:=2n^{i}_{y}+2n^{i}_{u}$. From \eqref{eqn-2}-\eqref{eqn-5}, the dynamics of $z_{i}\in\mathbb{R}^{n^{i}_{z}}$ can be written as
\begin{align}
\label{eqn-11}
\dot{z}_{i}&=f^{i}_{z}(z_{i}, x_{\p}, x_{\rf}, x^{i}_{\ct}, w_{\p}, w_{\rf}, w_{\ct}).
\end{align}
Stacking all $z_{i}$ leads to $z:=(z_{1}, \ldots, z_{N})\in\mathbb{R}^{n_{z}}$ with $n_{z}:=\sum^{N}_{i=1}n^{i}_{z}$, and we denote $\dot{z}=f_{z}:=(f^{1}_{z}, \ldots, f^{N}_{z})\in\mathbb{R}^{n_{z}}$. Because of the band-limited capacity of each network and the spatial locations of its sensors and actuators, all sensors and actuators of each network are grouped into $\ell_{i}\in\mathbb{N}_{+}$ nodes to access to the network, where $i\in\mathcal{N}$; see also \cite{Walsh2002stability, Nesic2004input}. Correspondingly, $z_{i}$ is partitioned into $\ell_{i}$ parts. For the $i$-th network, its sampling time sequence is denoted by $\{t^{i}_{j}: i\in\mathcal{N}, j\in\mathbb{N}_{+}\}$, which is strictly increasing. At the sampling time $t^{i}_{j}$, one and only one node is allowed to access to the $i$-th network, and this node is chosen by the time-scheduling protocol, which will be introduced in Section \ref{subsec-protocol}. For the $i$-th network, the sampling intervals are defined as $h^{i}_{j}:=t^{i}_{j+1}-t^{i}_{j}$, where $i\in\mathcal{N}$ and $j\in\mathbb{N}_{+}$. Since it takes time to compute and transmit the information, each agent may not receive the transmitted information instantaneously. Hence, there exist transmission delays $\tau^{i}_{j}\geq0$ such that the transmitted information is received at the arrival time $r^{i}_{j}=t^{i}_{j}+\tau^{i}_{j}$. For all networks, the following assumption is to bound sampling intervals and transmission delays.

\begin{assumption}
\label{asn-3}
For the $i$-th network with $i\in\mathcal{N}$, there exist constants $T_{i}\geq\Delta_{i}\geq0$ and $\varepsilon_{i}\in(0, T_{i})$ such that $\varepsilon_{i}\leq h^{i}_{j}\leq T_{i}$ and $0\leq\tau^{i}_{j}\leq\min\{\Delta_{i}, h^{i}_{j}\}$ hold for all $j\in\mathbb{N}_{+}$.
\end{assumption}

In Assumption \ref{asn-3}, $T_{i}>0$ is called the \emph{maximally allowable sampling period (MASP)} for the $i$-th network, $\Delta_{i}\geq0$ is called the \emph{maximally allowable delay (MAD)}, and $\varepsilon_{i}>0$ is the minimal interval of two successive transmissions. Both MASP and the MAD are the design parameters and will be upper bounded in Section \ref{subsec-trackanalyz}. $\varepsilon_{i}>0$ is determined by the hardware constraints \cite{Heemels2010networked}, and thus ensures the exclusion of Zeno phenomena. In the network-free case \cite{Yu2018explicit}, $\varepsilon_{i}\equiv0$ and $0<h^{i}_{j}\leq T_{i}$. If $\varepsilon_{i}=0$ in the networked case, the results derived in this paper are available for the periodic ETC case \cite{Wang2019periodic, Yu2018explicit}; otherwise, the ETM needs to be re-designed; see, e.g., \cite{Dolk2016output} for the continuous ETC case.

To reduce the transmission frequency, a local ETM is implemented for each network. That is, at each sampling time $t^{i}_{j}$, only when the event-triggered condition for the $i$-th network is satisfied can the sampled information be transmitted via the $i$-th network. Denote by $\hat{z}:=(\hat{y}_{\p}, \hat{y}_{\rf}, \hat{u}_{\ct}, \hat{u}_{\f})\in\mathbb{R}^{n_{z}}$ the received measurement after the transmission, and thus the control input received by the agents is $\hat{u}:=\hat{u}_{\ct}+\hat{u}_{\f}$. The network-induced errors are defined as $e_{\p}:=\hat{y}_{\p}-y_{\p}$, $e_{\rf}:=\hat{y}_{\rf}-y_{\rf}$, $e_{\ct}:=\hat{u}_{\ct}-u_{\ct}$ and $e_{\f}:=\hat{u}_{\f}-u_{\f}$. From $N$ networks, we denote $\hat{z}=(\hat{z}_{1}, \ldots, \hat{z}_{N})$ and $\mathbf{e}:=z-\hat{z}=(\mathbf{e}_{1}, \ldots, \mathbf{e}_{N})\in\mathbb{R}^{n_{z}}$.

In the arrival interval $[r^{i}_{j}, r^{i}_{j+1}]$, the received measurement $\hat{z}_{i}$ is assumed to be implemented via the zero-order hold (ZOH) mechanism, that is,
\begin{align}
\label{eqn-12}
\dot{\hat{z}}_{i}(t)&=0, \quad \forall t\in[r^{i}_{j}, r^{i}_{j+1}].
\end{align}
At the arrival time $r^{i}_{j}$, $j\in\mathbb{N}_{+}$, whether $\hat{z}_{i}$ is updated via the latest information depends on the local ETM at $t^{i}_{j}$. Here, we assume that the event-triggered condition for the $i$-th network is given by $\Gamma_{i}\geq0$, where the function $\Gamma_{i}: \mathbb{R}_{\geq0}\rightarrow\mathbb{R}$ will be designed explicitly in Section \ref{subsec-DETM}. $\Gamma_{i}\geq0$ implies that the sampled measurement needs to be transmitted, and $\hat{z}_{i}$ is updated with the latest measurement. That is, $\hat{z}_{i}$ is updated by
\begin{align}
\label{eqn-13}
&\hat{z}_{i}({r^{i}_{j}}^{+})=\left\{\begin{aligned}
&z_{i}(t^{i}_{j})+\mathbf{h}^{i}_{z}(\kappa_{i}(t^{i}_{j}), \mathbf{e}_{i}(t^{i}_{j})), &\Gamma_{i}(t^{i}_{j})\geq0, \\
&\hat{z}_{i}(r^{i}_{j}), &\Gamma_{i}(t^{i}_{j})<0,
\end{aligned}\right.
\end{align}
where $\kappa_{i}: \mathbb{R}_{\geq0}\rightarrow\mathbb{N}$ is a counter to record the number of the successful transmission events. That is, $\kappa_{i}({t^{i}_{j}}^{+})=\kappa_{i}(t^{i}_{j})+1$ if $\Gamma_{i}(t^{i}_{j})\geq0$, and $\kappa_{i}({t^{i}_{j}}^{+})=\kappa_{i}(t^{i}_{j})$ otherwise. $\mathbf{h}^{i}_{z}\in\mathbb{R}^{n_{z}}$ is the update function and depends on the protocol. Let $\mathbf{h}^{i}_{z}:=(\mathbf{h}^{i}_{\p}, \mathbf{h}^{i}_{\rf}, \mathbf{h}^{i}_{\ct}, \mathbf{h}^{i}_{\f})$, and then we can rewrite \eqref{eqn-13} as
\begin{align}
\label{eqn-14}
\hat{z}_{i}({r^{i}_{j}}^{+})&=(1-\Upsilon(\Gamma_{i}(t^{i}_{j})))\hat{z}_{i}(r^{i}_{j}) \nonumber \\
&\quad  +\Upsilon(\Gamma_{i}(t^{i}_{j}))[z_{i}(t^{i}_{j})+\mathbf{h}^{i}_{z}(\kappa_{i}(t^{i}_{j}), \mathbf{e}_{i}(t^{i}_{j}))],
\end{align}
where $\Upsilon: \mathbb{R}\rightarrow\{0, 1\}$ is defined as $\Upsilon(\Gamma_{i})=1$ if $\Gamma_{i}\geq0$ and $\Upsilon(\Gamma_{i})=0$ otherwise. From \eqref{eqn-14}, the error $\mathbf{e}_{i}$ is updated by
\begin{align*}
\mathbf{e}_{i}({r^{i}_{j}}^{+})&=\hat{z}_{i}({r^{i}_{j}}^{+})-z_{i}({r^{i}_{j}}^{+}) \\
&=\mathbf{e}_{i}(r^{i}_{j})+\Upsilon(\Gamma_{i}(t^{i}_{j}))[\mathbf{h}^{i}_{z}(\kappa_{i}(t^{i}_{j}), \mathbf{e}_{i}(t^{i}_{j}))-\mathbf{e}_{i}(t^{i}_{j})].
\end{align*}

\subsection{Time-Scheduling Protocols}
\label{subsec-protocol}

Since each network has $\ell_{i}$ nodes with $i\in\mathcal{N}$, the time-scheduling protocol is introduced to determine which node to access to the network. Similar to the analysis and the terminology in \cite{Heemels2010networked}, the function $\mathbf{h}^{i}_{z}(\kappa_{i}, \mathbf{e}_{i})$ in \eqref{eqn-14} is referred to as the \emph{protocol}. Based on $\ell_{i}$ nodes for the $i$-th network, $\mathbf{e}_{i}$ is partitioned into $\mathbf{e}_{i}=(\mathbf{e}^{i}_{1}, \ldots, \mathbf{e}^{i}_{\ell_{i}})$. If the $l_{i}$-th node is granted to access to the $i$-th network, where $l_{i}\in\{1, \ldots, \ell_{i}\}$, then the corresponding component $\mathbf{e}^{i}_{l_{i}}$ is updated and the other components are kept constant. In the literature \cite{Heemels2010networked, Nesic2004input}, many time-scheduling protocols can be modeled as $\mathbf{h}^{i}_{z}(\kappa_{i}, \mathbf{e}_{i})$, and two classes of commonly-used protocols are recalled.

The first protocol is the Round-Robin (RR) protocol, which is a periodic protocol \cite{Nesic2004input}. The period of the RR protocol is $\ell_{i}$, and each node has one and only chance to access to the $i$-th network in a period. The function $\mathbf{h}^{i}_{z}$ is given by
\begin{equation}
\label{eqn-15}
\mathbf{h}^{i}_{z}(\kappa_{i}, \mathbf{e}_{i}):=(I-\Psi_{i}(\kappa_{i}))\mathbf{e}_{i}(t^{i}_{j}),
\end{equation}
where, $\Psi_{i}(\kappa_{i})=\diag\{\Psi^{i}_{1}(\kappa_{i}), \ldots, \Psi^{i}_{\ell_{i}}(\kappa_{i})\}$ and $\Psi^{i}_{l_{i}}(\kappa_{i})\in\mathbb{R}^{n_{l_{i}}\times n_{l_{i}}}$, $\sum^{\ell_{i}}_{l_{i}=1}n_{l_{i}}=n^{i}_{\mathbf{e}}$. $\Psi^{i}_{l_{i}}(\kappa_{i})=I$ if $\kappa_{i}=l_{i}+\jmath\ell_{i}$ with $\jmath\in\mathbb{N}$ and $l_{i}\in\{1, \ldots, \ell_{i}\}$; otherwise, $\Psi^{i}_{l_{i}}(\kappa_{i})=0$.

The second protocol is Try-Once-Discard (TOD) protocol, which is a quadratic protocol \cite{Walsh2002stability}. For the TOD protocol, the node with a minimum index where the norm of the local network-induced error is the largest is allowed to access to the network. The function $\mathbf{h}^{i}_{z}$ is given by
\begin{equation}
\label{eqn-16}
\mathbf{h}^{i}_{z}(\kappa_{i}, \mathbf{e}_{i}):=(I-\Psi_{i}(\mathbf{e}_{i}))\mathbf{e}_{i}(t^{i}_{j}),
\end{equation}
where, $\Psi_{i}(\mathbf{e}_{i})=\diag\{\Psi^{i}_{1}(\mathbf{e}_{i}), \ldots, \Psi^{i}_{\ell_{i}}(\mathbf{e}_{i})\}$, and $\Psi^{i}_{l_{i}}(\mathbf{e}_{i})=I$ if $\min\{\arg\max_{1\leq k\leq \ell_{i}}|\mathbf{e}^{i}_{k}|\}=l_{i}$; otherwise, $\Psi^{i}_{l_{i}}(\mathbf{e}_{i})=0$.

\begin{remark}
\label{rmk-2}
Beside RR and TOD protocols, the TOD-tracking protocol was introduced in \cite{Postoyan2014tracking} by refining the TOD protocol. If $u^{i}_{\f}$ is generated by the controller \cite{Van2010tracking}, then $\mathbf{h}^{i}_{z}$ is associated to $(e^{i}_{\rf}, e^{i}_{\ct}+e^{i}_{\f})$ instead of $\mathbf{e}_{i}$. If $y^{i}_{\p}-y^{i}_{\rf}$ is transmitted via the network \cite{Ren2019tracking}, then $\mathbf{h}^{i}_{z}$ is related to $(e^{i}_{\p}-e^{i}_{\rf}, e^{i}_{\ct}-e^{i}_{\f})$.
\hfill $\square$
\end{remark}

\section{Development of Hybrid Model}
\label{sec-developandreformulate}

After the detailed analysis in previous section, we construct the hybrid model for the event-triggered tracking control of networked MAS in this section. To this end, our objective is to guarantee the convergence of $x_{\p}$ towards $x_{\rf}$ in the presence of ETMs and multiple networks. To measure the convergence of $x_{\p}$ towards $x_{\rf}$, define the tracking error $\eta:=x_{\p}-x_{\rf}\in\mathbb{R}^{n_{\p}}$, and the variable $e_{\ax}:=(e_{\eta}, e_{\ct}):=(e_{\p}-e_{\rf}, e_{\ct})\in\mathbb{R}^{n_{\ax}}$ with the network-induced errors $e_{\p}, e_{\rf}, e_{\ct}$ defined in Section \ref{sec-problemformation}, where $n_{\ax}=n_{\p}+n_{\ct}$. Combining all variables and analyses in Section \ref{sec-problemformation}, we derive the following impulsive model:
\begin{subequations}
\label{eqn-17}
\begin{align}
\label{eqn-17a}
&\left.\begin{aligned}
\dot{\eta}&=F_{\eta}(\delta, \eta, x_{\ct}, x_{\rf}, e_{\ax}, e_{\f}, e_{\rf}, w)\\
\dot{x}_{\rf}&=F_{\rf}(\delta, \eta, x_{\ct}, x_{\rf}, e_{\ax}, e_{\f}, e_{\rf}, w)\\
\dot{x}_{\ct}&=F_{\ct}(\delta, \eta, x_{\ct}, x_{\rf}, e_{\ax}, e_{\f}, e_{\rf}, w)\\
\dot{e}_{\ax}&=G_{\ax}(\delta, \eta, x_{\ct}, x_{\rf}, e_{\ax}, e_{\f}, e_{\rf}, w)\\
\dot{e}_{\rf}&=G_{\rf}(\delta, \eta, x_{\ct}, x_{\rf}, e_{\ax}, e_{\f}, e_{\rf}, w)\\
\dot{e}_{\f}&=G_{\f}(\delta, \eta, x_{\ct}, x_{\rf}, e_{\ax}, e_{\f}, e_{\rf}, w) \\
\end{aligned}\right\} \quad t^{i}\in[r^{i}_{j}, r^{i}_{j+1}], \\
\label{eqn-17b}
&\begin{aligned}
e^{i}_{\ax}({r^{i}_{j}}^{+})&=e^{i}_{\ax}(r^{i}_{j})+\Upsilon(\Gamma_{i}(t^{i}_{j}))[-e^{i}_{\ax}(t^{i}_{j})  \\
&\quad +h^{i}_{\ax}(\kappa_{i}(t^{i}_{j}), e^{i}_{\ax}(t^{i}_{j}), e^{i}_{\rf}(t^{i}_{j}), e^{i}_{\f}(t^{i}_{j}))],\\
e^{i}_{\rf}({r^{i}_{j}}^{+})&=e^{i}_{\rf}(r^{i}_{j})+\Upsilon(\Gamma_{i}(t^{i}_{j}))[-e^{i}_{\rf}(t^{i}_{j})  \\
&\quad +h^{i}_{\rf}(\kappa_{i}(t^{i}_{j}), e^{i}_{\ax}(t^{i}_{j}), e^{i}_{\rf}(t^{i}_{j}), e^{i}_{\f}(t^{i}_{j}))],\\
e^{i}_{\f}({r^{i}_{j}}^{+})&=e^{i}_{\f}(r^{i}_{j})+\Upsilon(\Gamma_{i}(t^{i}_{j}))[-e^{i}_{\f}(t^{i}_{j})  \\
&\quad +h^{i}_{\f}(\kappa_{i}(t^{i}_{j}), e^{i}_{\ax}(t^{i}_{j}), e^{i}_{\rf}(t^{i}_{j}), e^{i}_{\f}(t^{i}_{j}))],
\end{aligned}
\end{align}
\end{subequations}
where $h^{i}_{\ax}=(\mathbf{h}^{i}_{\p}-\mathbf{h}^{i}_{\rf}, \mathbf{h}^{i}_{\ct}), h^{i}_{\rf}=\mathbf{h}^{i}_{\rf}$ and $h^{i}_{\f}=\mathbf{h}^{i}_{\f}$ with $(\mathbf{h}^{i}_{\p}, \mathbf{h}^{i}_{\rf}, \mathbf{h}^{i}_{\ct}, \mathbf{h}^{i}_{\f})$ defined in Section \ref{sec-problemformation}. In addition, $\delta:=(\delta_{1}, \ldots, \delta_{N})\in\mathbb{R}^{N}$, and $\delta_{i}\in\mathbb{R}_{\geq0}$ is to model the `continuous' time of the $i$-th network with $\delta_{i}({r^{i}_{j}}^{+})=\delta_{i}(r^{i}_{j})$ and depends on $u^{i}_{\f}$ and/or its differential \cite{Postoyan2014tracking, Ren2019tracking}. All functions in \eqref{eqn-17a} are derived by detailed calculations and given in Appendix \ref{sec-appendixA}. Now, our objective is to derive reasonable  conditions and ETMs synchronously to guarantee ISS of the system \eqref{eqn-17} from $(e_{\rf}, e_{\f}, w)$ to $(\eta, e_{\ax})$. Here, $(e_{\rf}, e_{\f})$ is the network-induced errors, and may not be vanishing with the time line \cite{Van2010tracking, Postoyan2014tracking}.

\subsection{Hybrid Model of Networked MAS}
\label{subsec-hybrid}

To facilitate the analysis afterwards, the impulsive model \eqref{eqn-16} is further transformed into a formal hybrid model in the formalism of \cite{Goebel2012hybrid}. For the sake of convenience, define $x:=(\eta, x_{\rf}, x_{\ct})\in\mathbb{R}^{n_{x}}$ and $e:=(e_{\ax}, e_{\rf}, e_{\f})\in\mathbb{R}^{n_{e}}$ with $n_{x}=n_{\p}+n_{\ct}+n_{\rf}$ and $n_{e}=n_{\ax}+n_{y}+n_{u}$. Define $m:=(m_{1}, \ldots, m_{N})\in\mathbb{R}^{n_{e}}$ with $m_{i}:=h_{i}(\kappa_{i}, e_{i})-e_{i}\in\mathbb{R}^{n^{i}_{e}}$ storing the information for the update, where $e_{i}:=(e^{i}_{\ax}, e^{i}_{\rf}, e^{i}_{\f})$ and $h_{i}:=(h^{i}_{\ax}, h^{i}_{\rf}, h^{i}_{\f})$ are defined in \eqref{eqn-17}. Let $\kappa:=(\kappa_{1}, \ldots, \kappa_{N})\in\mathbb{R}^{N}$ with $\kappa_{i}\in\mathbb{N}$ defined in \eqref{eqn-13}; $\tau:=(\tau_{1}, \ldots, \tau_{N})\in\mathbb{R}^{N}$ with $\tau_{i}\in\mathbb{R}$ defined as a timer to keep track of both sampling intervals and transmission delays for the $i$-th network; $b:=(b_{1}, \ldots, b_{N})\in\mathbb{R}^{N}$, where $b_{i}\in\{0, 1\}$ is a logical variable to show whether the next event is a sampling event or an update event. That is, for the $i$-th network, $b_{i}=0$ means that the next event will be the sampling event, while $b_{i}=1$ means that the next event will be the update event. Denote $\mathfrak{X}:=(x, e, m, \delta, \tau, \kappa, b)\in\mathscr{R}:=\mathbb{R}^{n_{x}}\times\mathbb{R}^{n_{e}}\times\mathbb{R}^{n_{e}}\times\mathbb{R}^{N}\times\mathbb{R}^{N}\times\mathbb{R}^{N}\times\{0, 1\}^{N}$, and the hybrid system is developed below.
\begin{align}
\label{eqn-18}
\left\{\begin{aligned}
&\dot{\mathfrak{X}}=F(\mathfrak{X}, w), &\quad& \mathfrak{X}\in C,  \\
&\mathfrak{X}^{+}=G(\mathfrak{X}), &\quad& \mathfrak{X}\in D,
\end{aligned}\right.
\end{align}
where
\begin{align}
\label{eqn-19}
C&:=\bigcup^{N}_{i=1} C_{i}, \quad D:=\bigcup^{N}_{i=1}(D_{1i}\cup D_{2i}), \\
C_{i}&:=\{\mathfrak{X}\in\mathscr{R}: (b_{i}, \tau_{i})\in(\{0\}\times[0, T_{i}])\cup(\{1\}\times[0, \Delta_{i}])\}, \nonumber \\
D_{1i}&:=\{\mathfrak{X}\in\mathscr{R}: (b_{i}, \tau_{i})\in\{0\}\times[\varepsilon_{i}, T_{i}]\}, \nonumber \\
D_{2i}&:=\{\mathfrak{X}\in\mathscr{R}: (b_{i}, \tau_{i})\in\{1\}\times[0, \Delta_{i}]\}, \nonumber
\end{align}
with $T_{i}>0$ and $\Delta_{i}\geq0$ from Assumption \ref{asn-3}. The map $F$ is
\begin{align}
\label{eqn-20}
F(\mathfrak{X}, w)&:=(f(\delta, x, e, w), g(\delta, x, e, w), 0, \mathds{E}, \mathds{E}, 0, 0),
\end{align}
where $f:=(F_{\eta}, F_{\rf}, F_{\ct})$ and $g:=(G_{\ax}, G_{\rf}, G_{\f})$ are derived from \eqref{eqn-17a}. The mapping $G$ in \eqref{eqn-18} is defined as
\begin{align}
\label{eqn-21}
G(\mathfrak{X})&:=\left\{\begin{aligned}
&G_{1}(\mathfrak{X}), &\quad& \mathfrak{X}\in D_{1}, \\
&G_{2}(\mathfrak{X}), &\quad& \mathfrak{X}\in D_{2},
\end{aligned}\right.
\end{align}
with
\begin{align}
\label{eqn-22}
&\begin{aligned}
G_{1}(\mathfrak{X})&=\bigcup^{N}_{i=1}G_{1i}(\mathfrak{X}), \quad D_{1}=\bigcup^{N}_{i=1}D_{1i}, \\
G_{1i}(\mathfrak{X})&:=\left\{\begin{aligned}
&\begin{bmatrix}
x \\ e\\ \mathcal{M}_{1i}(x, e, m, \kappa) \\ \delta \\ \Lambda_{i}\tau \\ \kappa+\Upsilon(\Gamma_{i})(I-\Lambda_{i})\mathds{E} \\ b+(I-\Lambda_{i})\mathds{E}
\end{bmatrix}, &\quad& \mathfrak{X}\in D_{1i}, \\
&\varnothing, &\quad& \mathfrak{X}\notin D_{1i},
\end{aligned}\right.
\end{aligned}  \\
\label{eqn-23}
&\begin{aligned}
G_{2}(\mathfrak{X})&=\bigcup^{N}_{i=1}G_{2i}(\mathfrak{X}), \quad D_{2}=\bigcup^{N}_{i=1}D_{2i},\\
G_{2i}(\mathfrak{X})&:=\left\{\begin{aligned}
&\begin{bmatrix}
x \\ \mathcal{E}_{i}(x, e, m, \kappa) \\ \mathcal{M}_{2i}(x, e, m, \kappa) \\ \delta\\  \tau  \\ \kappa  \\  \Lambda_{i}b
\end{bmatrix}, &\quad& \mathfrak{X}\in D_{2i},  \\
&\varnothing, &\quad& \mathfrak{X}\notin D_{2i},
\end{aligned}\right.
\end{aligned}
\end{align}
where $\Lambda_{i}:=\diag\{\Lambda^{1}_{i}, \ldots, \Lambda^{N}_{i}\}\in\mathbb{R}^{N\times N}$ with $\Lambda^{k}_{i}=0$ if $k=i\in\mathcal{N}$ and $\Lambda^{k}_{i}=1$ otherwise, and
\begin{align*}
\mathcal{M}_{1i}(x, e, m, \kappa)&:=\Phi_{i}m+(I-\Phi_{i})M_{1i}(x, e, m, \kappa), \\
\mathcal{M}_{2i}(x, e, m, \kappa)&:=\Phi_{i}m+(I-\Phi_{i})M_{2i}(e, m), \\
\mathcal{E}_{i}(x, e, m, \kappa)&:=\Phi_{i}m+\Upsilon(\Gamma_{i})(I-\Phi_{i})E_{i}(e, m).
\end{align*}
Here, $\Phi_{i}:=\diag\{\Phi^{1}_{i}, \ldots, \Phi^{N}_{i}\}\in\mathbb{R}^{n_{e}\times n_{e}}, M_{1i}:=(M^{1}_{1i}, \ldots,$ $M^{N}_{1i})\in\mathbb{R}^{n_{e}}, M_{2i}:=(M^{1}_{2i}, \ldots, M^{N}_{2i})\in\mathbb{R}^{n_{e}}$ and $E_{i}:=(E^{1}_{i}, \ldots, E^{N}_{i})\in\mathbb{R}^{n_{e}}$. If $k=i$, then $\Phi^{k}_{i}=0$, $M^{k}_{1i}=(1-\Upsilon(\Gamma_{i}))m_{i}+\Upsilon(\Gamma_{i})(h_{i}(\kappa_{i}, e_{i})-e_{i}), M^{k}_{2i}=-e_{i}-m_{i}$ and $E^{k}_{i}=e_{i}+m_{i}$; otherwise, $\Phi^{k}_{i}=I$ and $M^{k}_{1i}=M^{k}_{2i}=E^{k}_{i}=0$.

For the hybrid model \eqref{eqn-18}, the sets $C$ and $D$ are closed. Since $f_{\p}, f_{\ct}, g_{\p}$ and $g_{\ct}$ are assumed to be continuous in Section \ref{subsec-trackingproblem}, $f$ and $g$ in \eqref{eqn-20} are continuous, and thus the flow map $F$ in \eqref{eqn-20} is continuous. The jump map $G$ in \eqref{eqn-21} is continuous and locally bounded from the continuity of $G_{1}$ in \eqref{eqn-22} and $G_{2}$ in \eqref{eqn-23}. As a result, we can verify easily that the hybrid model \eqref{eqn-18} satisfies the basic assumptions introduced in Section \ref{sec-preliminaries}.

\section{Tracking Performance Analysis}
\label{sec-mainresults}

In this section, the main results are established. We first present some necessary assumptions, then design the event-triggered condition for each network, and finally establish the convergence of the tracking error.

\subsection{Assumptions}
\label{subsec-assump}

Two types of assumptions are presented for the system \eqref{eqn-18}. The first type shows the properties of the $e_{i}$-subsystem in the flow and jumps, whereas the second type implies the stabilization property of the $x$-subsystem under the designed controller. We first present the first type of assumptions.

\begin{assumption}
\label{asn-4}
There exist a function $W_{i}: \mathbb{R}^{n^{i}_{e}}\times\mathbb{R}^{n^{i}_{e}}\times\mathbb{N}\times\{0, 1\}\rightarrow\mathbb{R}_{\geq0}$ which is locally Lipschitz in $(e_{i}, m_{i})$ for all $\kappa_{i}\in\mathbb{N}$ and $b_{i}\in\{0, 1\}$, $\alpha_{\jmath i}\in\mathcal{K}_{\infty}$, $\jmath\in\{1, \ldots, 6\}$, and $\lambda_{i}\in[0, 1)$ such that for all $(e_{i}, m_{i}, \kappa_{i}, b_{i})\in\mathbb{R}^{n^{i}_{e}}\times\mathbb{R}^{n^{i}_{e}}\times\mathbb{N}\times\{0, 1\}$,
\begin{align}
\label{eqn-24}
&\alpha_{1i}(|e^{i}_{\ax}|)\leq W_{i}(e_{i}, m_{i}, \kappa_{i}, b_{i})\leq\alpha_{2i}(|e_{i}|),\\
\label{eqn-25}
&W_{i}(e_{i}, h_{i}(\kappa_{i}, e_{i})-e_{i}, \kappa_{i}+1, 1) \nonumber \\
&\quad \leq\lambda_{i}W_{i}(e_{i}, m_{i}, \kappa_{i}, 0)+\alpha_{3i}(|e^{i}_{\f}|)+\alpha_{4i}(|e^{i}_{\rf}|), \\
\label{eqn-26}
&W_{i}(e_{i}, m_{i}, \kappa_{i}, 1)\leq W_{i}(e_{i}, m_{i}, \kappa_{i}, 0), \\
\label{eqn-27}
&W_{i}(e_{i}+m_{i}, -e_{i}-m_{i}, \kappa_{i}, 0)\nonumber \\
&\quad \leq W_{i}(e_{i}, m_{i}, \kappa_{i}, 1)+\alpha_{5i}(|e^{i}_{\f}|)+\alpha_{6i}(|e^{i}_{\rf}|).
\end{align}
\end{assumption}

\begin{assumption}
\label{asn-5}
There exist a continuous function $H_{ib_{i}}: \mathbb{R}^{n_{x}}\times\mathbb{R}^{n_{e}}\rightarrow\mathbb{R}_{>0}$, $\sigma_{1ib_{i}}, \sigma_{2ib_{i}}, \sigma_{3ib_{i}}\in\mathcal{K}_{\infty}$ and $L_{ib_{i}}\in\mathbb{R}_{\geq0}$ such that for all $(x, w, \kappa_{i}, b_{i})\in\mathbb{R}^{n_{x}}\times\mathbb{R}^{n_{w}}\times\mathbb{N}\times\{0, 1\}$ and almost all $(e_{i}, m_{i})\in\mathbb{R}^{n^{i}_{e}}\times\mathbb{R}^{n^{i}_{e}}$,
\begin{align}
\label{eqn-28}
&\left\langle\frac{\partial W_{i}(e_{i}, m_{i}, \kappa_{i}, b_{i})}{\partial e_{i}}, g_{i}(\delta, x, e, w)\right\rangle\leq L_{ib_{i}}W_{i}(e_{i}, m_{i}, \kappa_{i}, b_{i})  \nonumber \\
& +H_{ib_{i}}(x, e)+\sigma_{1ib_{i}}(|e^{i}_{\f}|)+\sigma_{2ib_{i}}(|e^{i}_{\rf}|)+\sigma_{3ib_{i}}(|w|).
\end{align}
\end{assumption}

Assumptions \ref{asn-4}-\ref{asn-5} are on the $e_{i}$-subsystem, whose properties are described via the function $W_{i}$. Assumption \ref{asn-4} is to estimate the jumps of $W_{i}$ at the discrete-time instants. Specifically, \eqref{eqn-25} is for the successful transmission jumps (i.e., $\Gamma_{i}\geq0$) at $t^{i}_{j}$, \eqref{eqn-26} is for the failure transmission jumps (i.e., $\Gamma_{i}<0$) at $t^{i}_{j}$, and \eqref{eqn-27} is for the update jumps at $r^{i}_{j}$. Assumption \ref{asn-5} is to estimate the derivative of $W_{i}$ in the continuous-time intervals, and the coupling is shown via the function $H_{ib_{i}}$. Since Assumptions \ref{asn-4}-\ref{asn-5} are applied to the $e_{i}$-subsystem, \eqref{eqn-25}-\eqref{eqn-27} hold with respect to the additional items $e^{i}_{\rf}$ and $e^{i}_{\f}$, which are parts of $e_{i}$ and treated as the internal disturbances caused by the network. Similar conditions have been considered in existing works \cite{Ren2019tracking, Postoyan2014tracking}, where however only a common communication network and TTC are studied. In addition, $\alpha_{3i}$ (or $\alpha_{4i}$) and $\alpha_{5i}$ (or $\alpha_{6i}$) in Assumption \ref{asn-4} can be the same. For instance, \eqref{eqn-25}-\eqref{eqn-27} hold with $\bar{\alpha}_{3i}(v)=\bar{\alpha}_{5i}(v):=\max\{\alpha_{3i}(v), \alpha_{5i}(v)\}$ and $\bar{\alpha}_{4i}(v)=\bar{\alpha}_{6i}(v):=\max\{\alpha_{4i}(v), \alpha_{6i}(v)\}$.

\begin{assumption}
\label{asn-6}
There exist a locally Lipschitz function $V: \mathbb{R}^{n_{x}}\rightarrow\mathbb{R}_{\geq0}$, $\alpha_{1V}, \alpha_{2V}, \zeta_{1ib_{i}}, \zeta_{2ib_{i}}, \zeta_{3ib_{i}}, \zeta_{4ib_{i}}, \zeta_{5ib_{i}}, \zeta_{6ib_{i}}\in\mathcal{K}_{\infty}$, and $\mu, \theta_{ib_{i}}, \gamma_{ib_{i}}>0, \bar{L}_{ib_{i}}\in\mathbb{R}$ such that
\begin{align}
\label{eqn-29}
&\alpha_{1V}(|\eta|)\leq V(x)\leq\alpha_{2V}(|x|), \quad \forall x\in\mathbb{R}^{n_{x}},
\end{align}
and for all $(e_{i}, m_{i}, \kappa_{i}, b_{i})\in\mathbb{R}^{n^{i}_{e}}\times\mathbb{R}^{n^{i}_{e}}\times\mathbb{N}\times\{0, 1\}$ and almost all $x\in\mathbb{R}^{n_{x}}$,
\begin{align}
\label{eqn-30}
&\langle\nabla V(x), f(\delta, x, e, w)\rangle\leq-\mu V(x)+\sum^{N}_{i=1}\left[-H^{2}_{ib_{i}}(x, e)\right.  \nonumber\\
&\quad +(\gamma^{2}_{ib_{i}}-\theta_{ib_{i}})W^{2}_{i}(e_{i}, m_{i}, \kappa_{i}, b_{i})-K_{ib_{i}}(x, e, m, w)  \nonumber\\
&\quad -\varphi_{ib_{i}}(z_{i})+\zeta_{1ib_{i}}(|e^{i}_{\f}|)+\zeta_{2ib_{i}}(|e^{i}_{\rf}|)+\zeta_{3ib_{i}}(|w|)], \\
\label{eqn-31}
&\langle\nabla\varphi_{ib_{i}}(z_{i}), f^{i}_{z}(\delta, x, e, w)\rangle\leq\bar{L}_{ib_{i}}\varphi_{i}(z_{i})+K_{ib_{i}}(x, e, m, w)  \nonumber \\
&\quad +H^{2}_{ib_{i}}(x, e)+\zeta_{4ib_{i}}(|e^{i}_{\f}|)+\zeta_{5ib_{i}}(|e^{i}_{\rf}|)+\zeta_{6ib_{i}}(|w|),
\end{align}
where $H_{ib_{i}}$ is defined in Assumption \ref{asn-4}, $\varphi_{ib_{i}}: \mathbb{R}^{n^{i}_{z}}\rightarrow\mathbb{R}_{\geq0}$ is a locally Lipschitz function with $\varphi_{ib_{i}}(0)=0$, and $K_{ib_{i}}: \mathbb{R}^{n_{x}}\times\mathbb{R}^{n_{e}}\times\mathbb{R}^{n_{e}}\times\mathbb{R}^{n_{w}}\rightarrow\mathbb{R}_{\geq0}$ is a continuous function.
\end{assumption}

Assumption \ref{asn-6} is on the $x$-subsystem, whose properties are described via the function $V$. Under the designed controller \eqref{eqn-9}-\eqref{eqn-10}, \eqref{eqn-29}-\eqref{eqn-30} imply that the $\eta$-subsystem satisfies the ISS-like property from $(\sum^{N}_{i=1}W_{i}, e_{\f}, e_{\rf}, w)$ to $\eta$, and the $\mathcal{L}_{2}$-stability property from $(\sum^{N}_{i=1}W_{i}, e_{\f}, e_{\rf}, w)$ to $\sum^{N}_{i=1}H_{ib_{i}}$. This assumption is reasonable due to the implementation of the emulation-based approach, where the controller is assumed to be known \emph{a priori} to ensure the tracking performance robustly in the network-free case. Hence, in the presence of the networks, $(\sum^{N}_{i=1}W_{i}, e_{\f}, e_{\rf}, w)$ is treated as a whole disturbance from the interior and exterior of the agents. Moreover, \eqref{eqn-31} provides the growth bound on the derivative of the function $\varphi_{ib_{i}}$ on the flow. Note that the information of multiple networks is not required in Assumption \ref{asn-6}, and that $\varphi_{ib_{i}}$ will be applied to design the ETMs.

\begin{remark}
\label{rmk-3}
Assumptions \ref{asn-4}-\ref{asn-6} depend on the existence of the functions $W_{i}$ and $V$, which were investigated in \cite{Wang2019periodic, Ren2019tracking, Postoyan2014tracking, Heemels2010networked} in the context of stabilization. In particular, Assumptions \ref{asn-4}-\ref{asn-6} are reformulated into linear matrix inequalities in the linear case \cite{Wang2019periodic}; $W_{i}$ and $V$ were constructed explicitly in \cite{Dolk2016output} for the event-triggered stabilization problems. On the other hand, in Assumptions \ref{asn-4}-\ref{asn-6}, the parameters $\lambda_{i}, L_{ib_{i}}, \bar{L}_{ib_{i}}$ are to facilitate the design and analysis afterwards. If the time-scheduling protocols are specified, then $\lambda_{i}$ can be computed explicitly; see \cite{Ren2019tracking, Nesic2004input, Postoyan2014tracking, Heemels2010networked}. $\bar{L}_{ib_{i}}$ is applied to design decentralized ETMs, and $L_{ib_{i}}$ is used to determine the MASPs and MADs. In \eqref{eqn-30}, the term $\mu V(x)$ can be relaxed into $\boldsymbol{\alpha}(V(x))$ with $\boldsymbol{\alpha}\in\mathcal{K}_{\infty}$ \cite{Wang2019periodic}, and the following analysis is still valid. The existence of these parameters has been extensively studied for different cases; see, e.g., \cite{Ren2019tracking, Dolk2016output, Nesic2004input, Wang2019periodic, Yu2018explicit, Walsh2002stability, Postoyan2014tracking, Heemels2010networked}.
\hfill $\square$
\end{remark}

\subsection{Decentralized Event-Triggered Mechanisms}
\label{subsec-DETM}

With Assumptions \ref{asn-4}-\ref{asn-6}, we next show how to design the ETM for each network. For this purpose, the function $\Gamma_{i}$ in \eqref{eqn-13} is defined as a mapping from $\mathbb{R}^{n^{i}_{z}}\times\mathbb{R}^{n^{i}_{e}}\times\mathbb{R}^{n^{i}_{e}}\times\mathbb{N}\times\{0, 1\}$ to $\mathbb{R}$, and given explicitly by
\begin{align}
\label{eqn-32}
\Gamma_{i}(z_{i}, e_{i}, m_{i}, \kappa_{i}, b_{i})&:=(1-2b_{i})\gamma_{ib_{i}}W^{2}_{i}(e_{i}, m_{i}, \kappa_{i}, b_{i}) \nonumber \\
&\quad -(1-b_{i})\rho_{i}\bar{\lambda}_{i}\varphi_{ib_{i}}(z_{i}),
\end{align}
where $W_{i}$ and $\varphi_{ib_{i}}$ are from in Assumptions \ref{asn-4} and \ref{asn-6}, respectively. $\rho_{i}\in[0, \bar{\rho}_{i})$, and
\begin{align}
\label{eqn-33}
\bar{\lambda}_{i}&:=\max\left\{\lambda_{i}, \frac{\rho_{i}\gamma_{i0}}{1-\rho_{i}\bar{L}_{i0}}\right\},\\
\label{eqn-34}
\bar{\rho}_{i}&:=\left\{\begin{aligned}
&1, &\quad& \bar{L}_{i0}\leq-\gamma_{i0},  \\
&\min\{1, (\bar{L}_{i0}+\gamma_{i0})^{-1}\}, &\quad& \bar{L}_{i0}>-\gamma_{i0},
\end{aligned}\right.
\end{align}
with $\lambda_{i}$ from Assumption \ref{asn-4} and $\gamma_{i0}, \bar{L}_{i0}$ from Assumption \ref{asn-6}.

With the function \eqref{eqn-32}, the event-triggered condition is $\Gamma_{i}(z_{i}, e_{i}, m_{i}, \kappa_{i}, b_{i})\geq0$. The proposed event-triggered condition is similar to those in \cite{Wang2019periodic, Tabuada2007event, Dimarogonas2011distributed} for the ETC in different contexts. Note that the function $\Gamma_{i}$ is related to the local measurements and thus only for the single network, which in turn leads to the decentralized ETC setting in this paper. One difference between \eqref{eqn-32} and the existing ones lies in the local logical variable $b_{i}$, which leads to two cases in \eqref{eqn-32}. Since the case $b_{i}=1$ implies that the update event will occur at the arrival instant, the ETM is not needed and $\Gamma_{i}(z_{i}, e_{i}, m_{i}, \kappa_{i}, 1)=-\gamma_{i1}W^{2}_{i}(e_{i}, m_{i}, \kappa_{i}, 1)<0$, which thus implies that the ETM will not be implemented in this case. In contrast, for the case $b_{i}=0$, the next event is the transmission event, and the ETM is implemented to determine whether the sampled measurement will be transmitted. Hence, $\Gamma_{i}(z_{i}, e_{i}, m_{i}, \kappa_{i}, 0)=\gamma_{i0}W^{2}_{i}(e_{i}, m_{i}, \kappa_{i}, 0)-\rho_{i}\bar{\lambda}_{i}\varphi_{i0}(z_{i})\geq0$ will be verified in this case. As a result, the parameters in \eqref{eqn-33}-\eqref{eqn-34} only depend on the case $b_{i}=0$, and all designed event-triggered conditions are consistent with the transmission setup and decentralized since only local information is involved in each event-triggered condition.

\begin{remark}
\label{rmk-4}
In \eqref{eqn-32}, if $\rho_{i}\equiv0$ for some $i\in\mathcal{N}$, then $\Gamma_{i}$ is always positive, and thus the proposed ETC is reduced to the TTC as in \cite{Ren2019tracking}, where $T_{i}$ is called the maximally allowable transmission interval. Since all networks are independent, both TTC and ETC can be combined by allowing that some networks perform the TTC while the others perform the ETC, which is a potential extension of this paper.
\hfill$\square$
\end{remark}

To establish the tradeoff between the MASP $T_{i}$ and the MAD $\Delta_{i}$, consider the following differential equation
\begin{align}
\label{eqn-35}
\dot{\phi}_{ib_{i}}&=-2L_{ib_{i}}\phi_{ib_{i}}-\gamma_{ib_{i}}[(1+\varrho_{ib_{i}})\phi^{2}_{ib_{i}}+1],
\end{align}
where $i\in\mathcal{N}$, $L_{ib_{i}}\geq0$ is given in Assumption \ref{asn-5}, and $\gamma_{ib_{i}}>0$ is given in Assumption \ref{asn-6}. In \eqref{eqn-35}, $\varrho_{ib_{i}}\in(0, \bar{\lambda}^{-2}_{i}\phi^{-2}_{ib_{i}}(0)-1)$, and thus the initial values $\phi_{ib_{i}}(0)\in(1, \bar{\lambda}^{-1}_{i})$, where $\bar{\lambda}_{i}$ is given in \eqref{eqn-33}. From Claim 1 in \cite{Carnevale2007lyapunov} and Claim 1 in \cite{Postoyan2014tracking}, the solutions to \eqref{eqn-35} are strictly decreasing as long as $\phi_{ib_{i}}\geq0$.

\subsection{Tracking Performance Analysis}
\label{subsec-trackanalyz}

Now we are ready to state the main results of this section.

\begin{theorem}
\label{thm-1}
Consider the system \eqref{eqn-18} and let Assumptions \ref{asn-1}-\ref{asn-6} hold. If the MASP $T_{i}$ and the MAD $\Delta_{i}$ satisfy
\begin{subequations}
\label{eqn-36}
\begin{align}
\label{eqn-36-1}
\gamma_{i0}\phi_{i0}(\tau_{i})&\geq(1+\varrho_{i1})\bar{\lambda}^{2}_{i}\gamma_{i1}\phi_{i1}(0), &\quad& \tau_{i}\in[0, T_{i}],\\
\label{eqn-36-2}
\gamma_{i1}\phi_{i1}(\tau_{i})&\geq(1+\varrho_{i0})\gamma_{i0}\phi_{i0}(\tau_{i}), &\quad& \tau_{i}\in[0, \Delta_{i}],
\end{align}
\end{subequations}
where $\phi_{ib_{i}}$ is the solution to \eqref{eqn-35} satisfying $\phi_{ib_{i}}(0), \phi_{ib_{i}}(T_{i})>0$, then the system \eqref{eqn-18} is ISS from $(e_{\rf}, e_{\f}, w)$ to $(\eta, e_{\ax})$. That is, there exist $\beta\in\mathcal{KLL}$ and $\varphi_{1}\in\mathcal{K}_{\infty}$ such that for all $(t, j)\in\mathbb{R}_{\geq0}\times\mathbb{N}$,
\begin{align}
\label{eqn-37}
|(\eta(t, j), e_{\ax}(t, j))|&\leq\beta(|\mathfrak{X}(0, 0)|, t, j)+\varphi_{1}(\|e_{\f}\|_{(t, j)}) \nonumber\\
&\quad +\varphi_{2}(\|e_{\rf}\|_{(t, j)})+\varphi_{3}(\|w\|_{(t, j)}).
\end{align}
\end{theorem}

The proof is presented in Appendix \ref{sec-appendixB}. In particular, a novel Lyapunov function is proposed to investigate the effects on the sampling, the designed ETMs and time delays on the tracking performance. Comparing with existing works \cite{Postoyan2014tracking, Ren2019tracking, Dolk2016output, Wang2019periodic} on NCS and \cite{Hong2006tracking, Cheng2016event} on MAS, the event-triggered tracking control problem is studied here for networked MAS under both decentralized ETMs and network constraints. Theorem \ref{thm-1} implies the convergence of the tracking error to a region around the origin, and the size of the convergence region depends on the network-induced error $(e_{\rf}, e_{\f})$ and the external disturbance $w$. If the external disturbance is not considered here, then the convergence region is only related to $(e_{\rf}, e_{\f})$. If the feedforward control inputs are transmitted directly to the agents and reference system, then $e_{\f}=0$ and $\varphi_{1}\equiv0$, and thus the convergence region can be further smaller.

\begin{remark}
\label{rmk-5}
From Theorem \ref{thm-1}, the conservatism is from the ISS gains $\varphi_{1}, \varphi_{2}$ in \eqref{eqn-37}. Since the ZOH mechanism is applied, the bounds on $\|e_{\f}\|_{(t, j)}, \|e_{\rf}\|_{(t, j)}$ can be obtained via a step-by-step sampling approach \cite{Van2010tracking}, and the upper bounds on $\varphi_{1}, \varphi_{2}$ can be derived. From \cite{Postoyan2014tracking, Ren2019tracking}, $\varphi_{1}, \varphi_{2}$ can be the functions of the MASP, the MAD and $\varepsilon_{i}$. Since $\varepsilon_{i}$ is from the hardware constraints of the networks, the effects of $\varepsilon_{i}$ are inevitable but can be limited by choosing appropriate networks, thereby leading to the lower bounds on $\varphi_{1}, \varphi_{2}$.
\hfill $\square$
\end{remark}

Next, the special case of a single network is addressed, that is, $N=1$ and only a common network is implemented for all agents. The following theorem shows how to ensure the tracking performance under the designed centralized ETM.

\begin{theorem}
\label{thm-2}
Consider the system \eqref{eqn-18} and let Assumptions \ref{asn-1}-\ref{asn-3} hold. If the following holds:
\begin{enumerate}[i)]
  \item for all $e\in\mathbb{R}^{n_{e}}\times\mathbb{R}^{l}\times\mathbb{R}^{n_{e}}\times\mathbb{R}^{l}\times\mathbb{N}\times\{0, 1\}$,
\begin{align}
\label{eqn-38}
&\alpha_{1W}(|e_{\ax}|)\leq W(e, m, \kappa, b)\leq\alpha_{2W}(|e|),\\
\label{eqn-39}
&W(e, h(\kappa, e)-e, \kappa+1, 1) \nonumber \\
&\quad \leq\lambda W(e, m, \kappa, 0)+\alpha_{3W}(|e_{\f}|)+\alpha_{4W}(|e_{\rf}|), \\
\label{eqn-40}
&W(e+m, -e-m, \kappa, 0)\nonumber \\
&\quad \leq W(e, m, \kappa, 1)+\alpha_{5W}(|e_{\f}|)+\alpha_{6W}(|e_{\rf}|).
\end{align}

\item for all $(x, \kappa)\in\mathbb{R}^{n_{x}}\times\mathbb{N}$ and almost all $e\in\mathbb{R}^{n_{e}}$,
\begin{align}
\label{eqn-41}
&\left\langle\frac{\partial W(e, m, \kappa, b)}{\partial e}, g_{e}(\delta, x, e, w)\right\rangle\leq L_{b}W(e, m, \kappa, b)  \nonumber \\
&\quad +H(x, e)+\sigma_{1b}(|e_{\f}|)+\sigma_{2b}(|e_{\rf}|)+\sigma_{3b}(|w|);
\end{align}

\item for all $(e, m, \kappa)\in\mathbb{R}^{n_{e}}\times\mathbb{R}^{n_{e}}\times\mathbb{N}$ and almost all $x\in\mathbb{R}^{n_{x}}$,
\begin{align}
\label{eqn-42}
&\langle\nabla V(x), f(\delta, x, e, w)\rangle\leq-\mu V(x)-H^{2}_{b}(x, e)  \nonumber\\
&\qquad\qquad +(\gamma^{2}_{b}-\theta)W^{2}(e, m, \kappa, b)+\zeta_{1b}(|e_{\f}|)  \nonumber\\
&\qquad \qquad +\zeta_{2b}(|e_{\rf}|)+\zeta_{3b}(|w|);
\end{align}

\item consider the following equation
\begin{align}
\label{eqn-43}
\dot{\phi}_{b}&=-2L_{b}\phi_{b}-\gamma_{b}[(1+\varrho_{b})\phi^{2}_{b}+1], \quad b\in\{0, 1\},
\end{align}
and the MASP $T$ and MAD $\Delta$ satisfy
\begin{subequations}
\label{eqn-44}
\begin{align}
\gamma_{0}\phi_{0}(\tau)&\geq(1+\varrho_{1})\bar{\lambda}^{2}\gamma_{1}\phi_{1}(0), &\ & \tau\in[0, T],\\
\gamma_{1}\phi_{1}(\tau)&\geq(1+\varrho_{0})\gamma_{0}\phi_{0}(\tau), &\ & \tau\in[0, \Delta],
\end{align}
\end{subequations}
\end{enumerate}
then the system \eqref{eqn-18} is ISS from $(e_{\rf}, e_{\f}, w)$ to $(\eta, e_{\ax})$ under the ETM designed below:
\begin{align}
\label{eqn-45}
\Gamma(x, e, m, \kappa, b)&=(1-2b)\gamma_{b}W^{2}(e, m, \kappa, b) \nonumber \\
&\quad -(1-b)\rho\bar{\lambda}V(x),
\end{align}
where $\bar{\lambda}:=\lambda\max\{1, \gamma_{0}\mu^{-1}\}$ and $\rho<\bar{\rho}=\min\{1, \mu\gamma^{-1}_{0}\}$.
\end{theorem}

The proof of Theorem \ref{thm-2} is given in Appendix \ref{sec-appendixB}. In Theorem \ref{thm-2}, the assumptions and the upper bounds in \eqref{eqn-44} are simplified and similar to those in \cite{Heemels2010networked, Ren2019tracking}. Although Theorem \ref{thm-2} is treated as a special case of Theorem \ref{thm-1}, \eqref{eqn-45} is based on the agent states and is different from \eqref{eqn-32} based on the agent outputs. In addition, Theorem \ref{thm-2} extends the results in \cite{Postoyan2014tracking, Heemels2010networked, Ren2019tracking} from the TTC case to the ETC case.

\begin{remark}
\label{rmk-6}
The derived conditions can be verified in different cases. For the linear case, these conditions can be transformed into linear matrix inequalities (LMIs) (\cite{Postoyan2014tracking, Wang2019periodic}) and checked by solving these LMIs. For the nonlinear case, the verification of these conditions depends on the considered system and the applied network. For instance, Assumption \ref{asn-4} involves the jumps of the function $W$. If the network protocols are specified, then the jumps of the network-induced errors can be derived explicitly, and the jumps of $W$ are established such that the corresponding parameters and functions can be determined \cite{Ren2019tracking, Postoyan2014tracking, Nesic2004input, Wang2019periodic, Walsh2002stability}. In Assumptions \ref{asn-5}-\ref{asn-6}, the derivatives of $W$ and $V$ are related to the considered system, and we present a relevant numerical example in Section \ref{subsec-example1} to show the satisfaction of Assumptions \ref{asn-5}-\ref{asn-6}.
\hfill $\square$
\end{remark}

\section{Event-Triggered Observer Design}
\label{sec-observer}

In this section, we apply the obtained results in the previous sections to the event-triggered observer design for networked MAS in the delay-free case. Consider the following MAS
\begin{align}
\label{eqn-46}
\dot{x}_{\p}=f_{\p}(x_{\p}, w), \quad  y_{\p}=g_{\p}(x_{\p}),
\end{align}
where $x_{\p}\in\mathbb{R}^{n_{\p}}$ is the system state, $w\in\mathbb{R}^{n_{w}}$ is the external disturbance, and $y_{\p}\in\mathbb{R}^{n_{y}}$ is the system output. The system \eqref{eqn-46} consists of $N$ agents with the following form
\begin{align}
\label{eqn-47}
\dot{x}^{i}_{\p}=f^{i}_{\p}(x_{\p}, w),\quad y^{i}_{\p}=g^{i}_{\p}(x^{i}_{\p}),
\end{align}
where $x_{\p}:=(x^{1}_{\p}, \ldots, x^{N}_{\p})\in\mathbb{R}^{n_{\p}}$, and $y_{\p}:=(y^{1}_{\p}, \ldots, y^{N}_{\p})\in\mathbb{R}^{n_{y}}$ with $n_{\p}:=\sum^{N}_{i=1}n^{i}_{\p}$ and $n_{y}:=\sum^{N}_{i=1}n^{i}_{y}$.

To design the distributed event-triggered observers for the MAS \eqref{eqn-45}, we consider the following two cases: the first case is that each observer only receives the information of the corresponding agent to estimate the state of this agent, and thus is called the \emph{decoupled observer design case}; the second case is that multiple agents are treated as a whole plant as \eqref{eqn-45} and all observers only receive partial information and thus need to be coupled to construct the plant state, which is called the \emph{coupled observer design case}.

\begin{figure}[!t]
\begin{center}
\begin{picture}(70, 90)
\put(-65, -13){\resizebox{70mm}{35mm}{\includegraphics[width=2.5in]{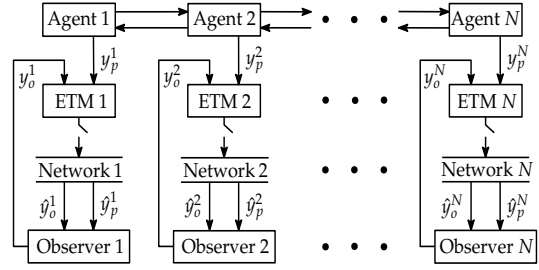}}}
\end{picture}
\end{center}
\caption{Configuration of the decoupled event-triggered observer design for networked MAS.}
\label{fig-2}
\end{figure}

\subsection{Decoupled Observer Design}
\label{subsec-decoupleobsever}

Assume that the observer is designed as
\begin{align}
\label{eqn-48}
\dot{x}_{\ob}=f_{\ob}(x_{\ob}, y_{\p}-y_{\ob}), \quad y_{\ob}=g_{\p}(x_{\ob}),
\end{align}
where $x_{\ob}\in\mathbb{R}^{n_{\p}}$ is the observer state, and $y_{\ob}\in\mathbb{R}^{n_{y}}$ is the output. Specifically, the observer for each agent is given by
\begin{align}
\label{eqn-49}
\dot{x}^{i}_{\ob}=f^{i}_{\ob}(x^{i}_{\ob}, y^{i}_{\p}-y^{i}_{\ob}),\quad y^{i}_{\ob}=g^{i}_{\p}(x^{i}_{\ob}).
\end{align}
The designed observers are distributed and decoupled; see Fig. \ref{fig-2} for the general system structure.

Here we aim to guarantee the desired estimation precision, that is, the convergence of $x_{\ob}$ towards $x_{\p}$, under the implementation of the designed observer \eqref{eqn-49} over multiple networks, which thus can be treated as a tracking problem with \eqref{eqn-46} being the reference system to be tracked by \eqref{eqn-48}.

\subsubsection{System Model}
Define the estimation error $\eta:=x_{\ob}-x_{\p}\in\mathbb{R}^{n_{\p}}$, $e_{\p}=\hat{y}_{\p}-y_{\p}$, $e_{\ob}=\hat{y}_{\ob}-y_{\ob}$, and $e_{\ax}=e_{\ob}-e_{\p}$. Define $x:=(\eta, x_{\p})\in\mathbb{R}^{n_{x}}$ and $e:=(e_{\ax}, e_{\p})\in\mathbb{R}^{n_{e}}$, where $e_{\ax}:=(e^{1}_{\ax}, \ldots, e^{N}_{\ax})$,  $e_{\p}:=(e^{1}_{\p}, \ldots, e^{N}_{\p})$, $n_{x}=2n_{\p}$ and $n_{e}=2n_{y}$. As a result, the impulsive model is given by
\begin{align}
\label{eqn-50}
&\left.\begin{aligned}
\dot{\eta}&=F_{\eta}(\eta, x_{\p}, e_{\ax}, w) \\
\dot{x}_{\p}&=F_{\p}(x_{\p}, w) \\
\dot{e}_{\ax}&=G_{\ax}(\eta, x_{\p}, e_{\ax}, w) \\
\dot{e}_{\p}&=G_{\p}(\eta, x_{\p}, e_{\ax}, w)
\end{aligned}\right\}  \quad t^{i}\in[t^{i}_{j}, t^{i}_{j+1}], \\
\label{eqn-51}
&\begin{aligned}
e^{i}_{\ax}({t^{i}_{j}}^{+})&=(1-\Upsilon(\Gamma_{i}(t^{i}_{j})))e^{i}_{\ax}(t^{i}_{j})  \\
&\quad+\Upsilon(\Gamma_{i}(t^{i}_{j}))h^{i}_{\ax}(\kappa_{i}(t^{i}_{j}), e^{i}_{\ax}(t^{i}_{j}), e^{i}_{\p}(t^{i}_{j})),  \\
e^{i}_{\p}({t^{i}_{j}}^{+})&=(1-\Upsilon(\Gamma_{i}(t^{i}_{j})))e^{i}_{\p}(t^{i}_{j})  \\
&\quad +\Upsilon(\Gamma_{i}(t^{i}_{j}))h^{i}_{\p}(\kappa_{i}(t^{i}_{j}), e^{i}_{\ax}(t^{i}_{j}), e^{i}_{\p}(t^{i}_{j})),
\end{aligned}
\end{align}
with $F_{\eta}(\eta, x_{\p}, e_{\ax}, w)=f_{\ob}(\eta+x_{\p}, g_{\p}(x_{\p})-g_{\ob}(\eta+x_{\p})-e_{\ax})-f_{\p}(x_{\p}, w), F_{\p}(x_{\p}, w)=f_{\p}(x_{\p}, w), G_{\ax}(\eta, x_{\p}, e_{\ax}, w)=-\langle\nabla g_{\p}(x_{\ob}), f_{\ob}(\eta+x_{\p}, g_{\p}(x_{\p})-g_{\ob}(\eta+x_{\p})-e_{\ax})\rangle+\langle\nabla g_{\p}(x_{\p}), f_{\p}(x_{\p}, w)\rangle$ and $G_{\p}(\eta, x_{\p}, e_{\ax}, w)=-\langle\nabla g_{\p}(x_{\p}), f_{\p}(x_{\p}, w)\rangle$. In \eqref{eqn-51}, the function $\Gamma_{i}$ is to be designed for the ETM afterwards. All functions in \eqref{eqn-50} are assumed to be continuous. Following the same fashion in Section \ref{subsec-hybrid}, the impulsive model \eqref{eqn-50}-\eqref{eqn-51} can be further reformulated as a formal hybrid model as \eqref{eqn-18}.

\subsubsection{Observer Design}
Since the controller and time delays are not considered here, $u_{\ct}, u_{\f}$ do not exist and $\delta, m, b$ are not needed. Hence, $e_{\f}\equiv0, z_{i}:=(y^{i}_{\p}, y^{i}_{\ob}), e^{i}_{\rf}=e^{i}_{\p}$ and $e_{i}:=(e^{i}_{\ax}, e^{i}_{\p})$. The following assumption is made for the $e_{i}$-subsystem, and is similar to Assumptions \ref{asn-4}-\ref{asn-5}.

\begin{assumption}
\label{asn-7}
There exist a function $W_{i}: \mathbb{N}\times\mathbb{R}^{n^{i}_{e}}\rightarrow\mathbb{R}_{\geq0}$ which is locally Lipschitz in $e_{i}$, a continuous function $H_{i}: \mathbb{R}^{n_{x}}\times\mathbb{R}^{n_{e}}\rightarrow\mathbb{R}_{>0}$, $\alpha_{1i}, \alpha_{2i}, \alpha_{3i}, \sigma_{1i}, \sigma_{2i}\in\mathcal{K}_{\infty}$ and $\lambda_{i}\in[0, 1), L_{i}\geq0$ such that for all $(\kappa_{i}, x, w)\in\mathbb{N}\times\mathbb{R}^{n_{x}}\times\mathbb{R}^{n_{w}}$,
\begin{itemize}
  \item for all $e_{i}\in\mathbb{R}^{n^{i}_{e}}$, $\alpha_{1i}(|e^{i}_{\ax}|)\leq W_{i}(\kappa_{i}, e_{i})\leq\alpha_{2i}(|e_{i}|)$, and $W_{i}(\kappa_{i}+1, h_{i}(\kappa_{i}, e_{i}))\leq\lambda_{i}W_{i}(\kappa_{i}, e_{i})+\alpha_{3i}(|e^{i}_{\p}|)$;
  \item for almost all $e_{i}\in\mathbb{R}^{n^{i}_{e}}$, $\langle\frac{\partial W_{i}(\kappa_{i}, e_{i})}{\partial e_{i}}, g_{i}(x, e, w)\rangle\leq L_{i}W_{i}(\kappa_{i}, e_{i})+H_{i}(x, e)+\sigma_{1i}(|e^{i}_{\p}|)+\sigma_{2i}(|w|)$.
\end{itemize}
\end{assumption}

Similarly to Section \ref{subsec-DETM}, the function $\Gamma_{i}: \mathbb{R}^{n^{i}_{z}}\times\mathbb{R}^{n^{i}_{e}}\times\mathbb{N}\rightarrow\mathbb{R}$ is defined explicitly below
\begin{align}
\label{eqn-52}
\Gamma_{i}(z_{i}, e_{i}, \kappa_{i})&=\gamma_{i}W^{2}_{i}(\kappa_{i}, e_{i})-\rho_{i}\bar{\lambda}_{i}\varphi_{i}(z_{i}),
\end{align}
where $W_{i}$ is defined in Assumption \ref{asn-7}, $\varphi_{i}$ is the same as $\varphi_{i0}$ in Assumption \ref{asn-6}, and $\rho_{i}, \bar{\lambda}_{i}$ satisfy \eqref{eqn-33}-\eqref{eqn-34}. With the event-triggered condition $\Gamma_{i}\geq0$, we next provide the bound on the MASP $T_{i}$. For any $i\in\mathcal{N}$, consider the differential equation
\begin{align}
\label{eqn-53}
\dot{\phi}_{i}&=-2L_{i}\phi_{i}-\gamma_{i}[(1+\varrho_{i})\phi^{2}_{i}+1],
\end{align}
where $L_{i}$ and $\gamma_{i}$ are given in Assumptions \ref{asn-6}-\ref{asn-7}, respectively; $\phi_{i}(0)\in(1, \bar{\lambda}^{-1}_{i})$, and $\varrho_{i}\in(0, \bar{\lambda}^{-2}_{i}\phi^{-2}_{i}(0)-1)$. In the following, the convergence of the estimation error can be justified, which is stated in the following proposition.

\begin{proposition}
\label{prop-1}
For the system \eqref{eqn-50}-\eqref{eqn-51}, let Assumptions \ref{asn-1}-\ref{asn-3} and \ref{asn-7} hold. Let Assumption \ref{asn-6} hold with $\zeta_{1i}=\zeta_{4i}\equiv0$.
If the MASP $T_{i}$ satisfies $\phi_{i}(T_{i})>0$ with $\phi_{i}$ being the solution to \eqref{eqn-53}, then the system \eqref{eqn-50}-\eqref{eqn-51} is ISS from $(e_{\p}, w)$ to $(\eta, e_{\ax})$.
\end{proposition}

The proof follows the similar fashion as that of Theorem \ref{thm-1}, and thus is omitted here. In addition, we can derive the event-triggered observer design for the centralized case by adjusting Assumption \ref{asn-7} and the conditions \eqref{eqn-52}-\eqref{eqn-53} slightly.

\subsection{Coupled Observer Design}
\label{subsec-coupleobsever}

\begin{figure}[!t]
\begin{center}
\begin{picture}(60, 120)
\put(-60, -12){\resizebox{60mm}{45mm}{\includegraphics[width=2.5in]{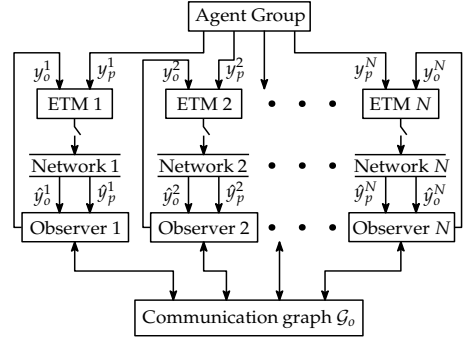}}}
\end{picture}
\end{center}
\caption{Framework for the coupled distributed event-triggered state estimation for networked MAS.}
\label{fig-3}
\end{figure}

In the coupled observer design case, each observer only receives partial information, and thus all observers need to exchange their information based on a pre-specified communication (di)graph \cite{Li2017robust, Park2016design}. Assume that in the network-free case, the distributed observers for \eqref{eqn-46} are designed as
\begin{align}
\label{eqn-54}
\begin{aligned}
\dot{x}^{i}_{\ob}&=f^{i}_{\p}(x^{i}_{\ob}, 0)+\mathfrak{F}_{i}(\vartheta_{i}),\quad y^{i}_{\ob}=g^{i}_{\p}(x^{i}_{\ob}), \\
\dot{\vartheta}_{i}&=\mathfrak{h}_{i}(y^{i}_{\ob}, \vartheta_{i}).
\end{aligned}
\end{align}
If the observer receives the latest information, the update is
\begin{align}
\label{eqn-55}
x^{i+}_{\ob}&=x^{i}_{\ob}, \quad \vartheta^{+}_{i}=\Theta_{i}(x_{\ob}, y_{\ob}, y_{\p}),
\end{align}
where $x_{\ob}=(x^{1}_{\ob}, \ldots, x^{N}_{\ob})\in\mathbb{R}^{n_{\p}N}$ and $y_{\ob}=(y^{1}_{\ob}, \ldots, y^{N}_{\ob})\in\mathbb{R}^{n_{y}N}$. In \eqref{eqn-54}-\eqref{eqn-55}, $\vartheta_{i}\in\mathbb{R}^{n^{i}_{\ob}}$ is an auxiliary variable to store the estimation errors of the $i$-th observer and its neighbor observers. $\mathfrak{F}_{i}$ is a continuous mapping from $\mathbb{R}^{n^{i}_{\ob}}$ to $\mathbb{R}^{n^{i}_{\ob}}$. The function $\Theta_{i}$ determines the update of the variable $\vartheta_{i}$. Although $\Theta_{i}$ is written as a function of the overall vectors $x_{\ob}, y_{\ob}, y_{\p}$, it depends only on the $i$-th observer and its neighbors. Fig. \ref{fig-3} shows the general structure for the event-triggered observer design in the coupled case. Here our objective is to guarantee the convergence of $x^{i}_{\ob}$ towards $x_{\p}$ under multiple networks and ETMs, which is treated as another tracking problem with \eqref{eqn-46} being the reference system to be tracked by \eqref{eqn-54}.

\subsubsection{System Model}
Define the estimation error $\eta_{i}:=x^{i}_{\ob}-x_{\p}\in\mathbb{R}^{n_{\p}}$, $e^{i}_{\p}=\hat{y}^{i}_{\p}-y^{i}_{\p}$, $e^{i}_{\ob}=\hat{y}^{i}_{\ob}-y^{i}_{\ob}$, $e^{i}_{\ax}=e^{i}_{\ob}-e^{i}_{\p}$, and $\psi_{i}=\Theta_{i}(x_{\ob}, y_{\ob}, y_{\p})-\vartheta_{i}$. Denote $\eta:=(\eta_{i}, \ldots, \eta_{N})\in\mathbb{R}^{n_{\p}N}$, and $e_{\p}:=(e^{1}_{\p}, \ldots, e^{N}_{\p})\in\mathbb{R}^{n_{\p}N}$. Hence, the impulsive model for the $i$-th observer is given by
\begin{align}
\label{eqn-56}
&\left.\begin{aligned}
\dot{\eta}_{i}&=F^{i}_{\eta}(\eta, x_{\p}, \psi_{i}, w) \\
\dot{x}_{\p}&=F_{\p}(x_{\p}, w)=f_{\p}(x_{\p}, w) \\
\dot{e}^{i}_{\ax}&=G^{i}_{\ax}(\eta_{i}, x_{\p}, \psi_{i}, w) \\
\dot{e}^{i}_{\p}&=G^{i}_{\p}(x_{\p}, w) \\
\dot{\psi}_{i}&=G^{i}_{\psi}(\eta, x_{\p}, \psi, e_{\ax}, e_{\p}, w)
\end{aligned}\right\}  \ t^{i}\in[t^{i}_{j}, t^{i}_{j+1}], \\
\label{eqn-57}
&\left.\begin{aligned}
e^{i}_{\ax}(t^{+})&=\Upsilon(\Gamma_{i}(t))h^{i}_{\ax}(\kappa_{i}(t), e^{i}_{\ax}(t), e^{i}_{\p}(t), \psi_{i}(t)) \\
&\quad +(1-\Upsilon(\Gamma_{i}(t)))e^{i}_{\ax}(t)   \\
e^{i}_{\p}(t^{+})&=\Upsilon(\Gamma_{i}(t))h^{i}_{\p}(\kappa_{i}(t), e^{i}_{\ax}(t), e^{i}_{\p}(t), \psi_{i}(t))  \\
&\quad +(1-\Upsilon(\Gamma_{i}(t)))e^{i}_{\p}(t)  \\
\psi_{i}(t^{+})&=(1-\Upsilon(\Gamma_{i}(t)))\psi_{i}(t)
\end{aligned}\right\}  \ t=t^{i}_{j},
\end{align}
with $G^{i}_{\p}(x_{\p}, w)=-\langle\nabla g^{i}_{\p}(x_{\p}), f_{\p}(x_{\p}, w)\rangle$ and
\begin{align*}
&F^{i}_{\eta}(\eta_{i}, x_{\p}, \psi_{i}, w)=f^{i}_{\p}(\eta_{i}+x_{\p}, 0)-f_{\p}(x_{\p}, w) \\
&\quad+\mathfrak{F}_{i}(\Theta_{i}(\eta, x_{\p}, e_{\p}, e_{\ax})-\psi_{i}), \\
&G^{i}_{\ax}(\eta_{i}, x_{\p}, \psi_{i}, w)=\langle\nabla g_{\p}(x_{\p}), F_{\p}(x_{\p}, w)\rangle \\
&\quad -\langle\nabla g^{i}_{\p}(x^{i}_{\ob}), f^{i}_{\p}(\eta_{i}+x_{\p}, 0)+\mathfrak{F}_{i}(\Theta_{i}(\eta, x_{\p}, e_{\p}, e_{\ax})-\psi_{i})\rangle, \\
&G^{i}_{\psi}(\eta, x_{\p}, \psi, e_{\p}, e_{\ax}, w)=\left\langle\frac{\partial\Theta_{i}(\eta, x_{\p}, e_{\p}, e_{\ax})}{\partial x_{\p}}, F_{\p}(x_{\p}, w)\right\rangle \\
&\quad +\sum_{j\in\mathcal{N}_{i}\cup\{i\}}\left\langle\frac{\Theta_{i}(\eta, x_{\p}, e_{\p}, e_{\ax})}{\partial\eta_{j}}, F^{j}_{\eta}(\eta_{j}, x_{\p}, \psi_{j}, w)\right\rangle \\
&\quad +\left\langle\frac{\partial\Theta_{i}(\eta, x_{\p}, e_{\p}, e_{\ax})}{\partial e^{i}_{\p}}, G^{i}_{\p}(x_{\p}, w)\right\rangle \\
&\quad +\sum_{j\in\mathcal{N}_{i}\cup\{i\}}\left\langle\frac{\Theta_{i}(\eta, x_{\p}, e_{\p}, e_{\ax})}{\partial e^{j}_{\ax}}, G^{j}_{\ax}(\eta_{j}, x_{\p}, \psi_{j}, w)\right\rangle \\
&\quad -\mathfrak{h}_{i}(g^{i}_{\p}(\eta_{i}+x_{\p})+e^{i}_{\p}-e^{i}_{\ax}, \Theta_{i}(\eta, x_{\p}, e_{\p}, e_{\ax})-\psi_{i}),
\end{align*}
where $\Theta_{i}(\eta, x_{\p}, e_{\p}, e_{\ax})$ is a rephrase of $\Theta_{i}(x_{\ob}, y_{\ob}, y_{\p})$ in \eqref{eqn-55}.

Different from the overall model \eqref{eqn-50}-\eqref{eqn-51} in the decoupled case, the model \eqref{eqn-56}-\eqref{eqn-57} is only for the single observer, and the coupling is shown in the function $g^{i}_{\psi}$. Define $x_{i}:=(\eta_{i}, x_{\p})$ and $e_{i}:=(e^{i}_{\p}, e^{i}_{\ax}, \psi_{i})$, and \eqref{eqn-56}-\eqref{eqn-57} can be written as
\begin{align}
\label{eqn-58}
\left.\begin{aligned}
\dot{x}_{i}&=f_{i}(x, e, w) \\
\dot{e}_{i}&=g_{i}(x, e, w)
\end{aligned}\right\}&  \ t\in[t^{i}_{j}, t^{i}_{j+1}],  \\
\label{eqn-59}
\left.\begin{aligned}
x_{i}(t^{+})&=x_{i}(t) \\
e_{i}(t^{+})&=\Upsilon(\Gamma_{i}(t))h_{i}(\kappa_{i}(t), e_{i}(t)) \\
&\quad +(1-\Upsilon(\Gamma_{i}(t)))e_{i}(t)
\end{aligned}\right\}&\   t=t^{i}_{j},
\end{align}
where $h_{i}:=(h^{i}_{\ax}, h^{i}_{\p}, 0)$. Let $z_{i}=(y_{\p}, y^{i}_{\ob})$, and $\dot{z}_{i}=(\langle\partial g_{\p}/\partial x_{\p}, F_{\p}(x_{\p}, w)\rangle, \langle\partial g^{i}_{\p}/\partial x^{i}_{\ob}, F^{i}_{\eta}(\eta, x_{\p}, \psi_{i}, w)+F_{\p}(x_{\p}, w)\rangle)=:f^{i}_{z}(x, e, w)$.

\subsubsection{Observer Design}
To ensure the convergence of the estimation error, the following assumption is imposed, which differs from Assumption \ref{asn-6} by allowing each observer to possess a Lyapunov function $V_{i}$ with $i\in\mathcal{N}$.

\begin{assumption}
\label{asn-8}
For each $i\in\mathcal{N}$, there exist a locally Lipschitz function $V_{i}: \mathbb{R}^{n^{i}_{x}}\rightarrow\mathbb{R}_{\geq0}$, $\bar{\alpha}_{i}, \underline{\alpha}_{i}, \zeta_{1i}, \zeta_{2i}, \zeta_{3i}, \zeta_{4i}\in\mathcal{K}_{\infty}$, and $\mu, \theta_{i}, \gamma_{i}>0, \bar{L}_{i}\in\mathbb{R}$, such that for all $(e_{i}, \kappa_{i})\in\mathbb{R}^{n^{i}_{e}}\times\mathbb{N}$,
\begin{itemize}
  \item for all $x\in\mathbb{R}^{n_{x}}$, $\overbar{\alpha}_{i}(|x_{i}|)\leq V_{i}(x_{i})\leq\underline{\alpha}_{i}(|x_{i}|)$,
  \item for almost all $x\in\mathbb{R}^{n_{x}}$, $\langle\nabla V_{i}(x_{i}), f_{i}(x, e, w)\rangle\leq-\mu_{i}V_{i}(x_{i})+(\gamma^{2}_{i}-\theta_{i})W^{2}_{i}(\kappa_{i}, e_{i})-H^{2}_{i}(x, e)-K_{i}(x, e, w)-\varphi_{i}(z_{i})+\zeta_{1i}(|e^{i}_{\p}|)+\zeta_{2i}(|w|)$, and $\langle\nabla\varphi_{i}(z_{i}), f^{i}_{z}(x, e, w)\rangle\leq\bar{L}_{i}\varphi_{i}(z_{i})+K_{i}(x, e, w)+H^{2}_{i}(x, e)+\zeta_{3i}(|e^{i}_{\p}|)+\zeta_{4i}(|w|)$, where $H_{i}$ is given in Assumption \ref{asn-7}, $\varphi_{i}$ is a locally Lipschitz function with $\varphi_{i}(0)=0$, and $K_{i}$ is a continuous non-negative function.
\end{itemize}
\end{assumption}

From Assumption \ref{asn-8}, Assumption \ref{asn-6} holds with $V(x):=\sum^{N}_{i=1}V_{i}(x_{i})$ and $\mu:=\min_{i\in\mathcal{N}}\{\mu_{i}\}$. In addition, we define the function $\Gamma_{i}$ as in \eqref{eqn-52}, and bound the MASP $T_{i}$ via \eqref{eqn-53}. In this way, we can ensure the convergence of the estimation error, which is stated in the following theorem.

\begin{proposition}
\label{prop-2}
Consider all subsystems of \eqref{eqn-58}-\eqref{eqn-59} and let Assumptions \ref{asn-1}-\ref{asn-3} and \ref{asn-7}-\ref{asn-8} hold. If the MASP $T_{i}$ satisfies $\phi_{i}(T_{i})>0$ with $\phi_{i}$ being the solution to \eqref{eqn-53}, then the whole system is ISS from $(e_{\p}, w)$ to $(\eta, e_{\ax})$.
\end{proposition}

The proof of Proposition \ref{prop-2} is similar to that of Theorem \ref{thm-1} and omitted here. The conditions derived in this section can be verified and the discussion in Remark \ref{rmk-6} is relevant here.

\section{Numerical Examples}
\label{sec-illustration}

\subsection{Tracking Control for Cooperative Manipulation}
\label{subsec-example1}

Consider two connected single-link robot arms, whose dynamics are presented as ($i=1, 2$)
\begin{align}
\label{eqn-60}
\begin{aligned}
\dot{\mathfrak{q}}^{i1}_{\p}&=\mathfrak{q}^{i2}_{\p}, \\
\dot{\mathfrak{q}}^{i2}_{\p}&=-a_{i}\sin\mathfrak{q}^{i1}_{\p}+\sum^{2}_{j=1}b_{ij}(\mathfrak{q}^{1j}_{\p}-\mathfrak{q}^{2j}_{\p})+c_{i}u_{i},
\end{aligned}
\end{align}
where $\mathfrak{q}^{i}_{\p}:=(\mathfrak{q}^{i1}_{\p}, \mathfrak{q}^{i2}_{\p})\in\mathbb{R}^{2}$ with the configuration coordinate $\mathfrak{q}^{i1}_{\p}$ and the velocity $\mathfrak{q}^{i2}_{\p}$, both of which are measurable, $u_{i}\in\mathbb{R}$ is the input torque, and $a_{i}, c_{i}>0, b_{ij}\in\mathbb{R}$ are certain constants. The references are given by
\begin{align}
\label{eqn-61}
\begin{aligned}
\dot{\mathfrak{q}}^{i1}_{\rf}&=\mathfrak{q}^{i2}_{\rf}, \\
\dot{\mathfrak{q}}^{i2}_{\rf}&=-a_{i}\sin\mathfrak{q}^{i1}_{\rf}+\sum^{2}_{j=1}b_{ij}(\mathfrak{q}^{1j}_{\rf}-\mathfrak{q}^{2j}_{\rf})+c_{i}u^{i}_{\f},
\end{aligned}
\end{align}
where $\mathfrak{q}^{i}_{\rf}:=(\mathfrak{q}^{i1}_{\rf}, \mathfrak{q}^{i2}_{\rf})\in\mathbb{R}^{2}$ are the measurable reference state, and $u^{i}_{\f}=c_{i}\sin(5t)$ is the feedforward input. In the network-free case, the feedback controller is designed as $u^{i}_{\ct}=-c^{-1}_{i}[a_{i}(\sin(\mathfrak{q}^{i1}_{\p})-\sin(\mathfrak{q}^{i1}_{\rf}))-(\mathfrak{q}^{i1}_{\p}
-\mathfrak{q}^{i1}_{\rf})-(\mathfrak{q}^{i2}_{\p}-\mathfrak{q}^{i2}_{\rf})]$ such that the tracking error is asymptotically stable.

\begin{figure}[!t]
\begin{center}
\begin{picture}(80,105)
\put(-80,-25){\resizebox{80mm}{50mm}{\includegraphics[width=2.5in]{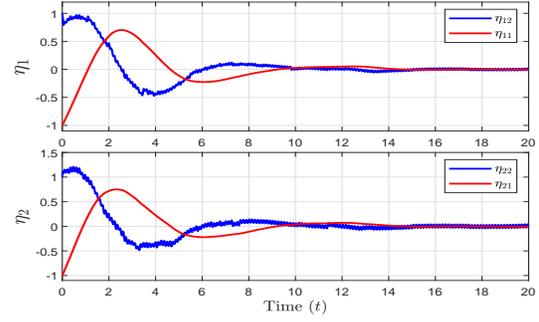}}}
\end{picture}
\end{center}
\caption{Tracking errors under the RR protocol case and the ETMs \eqref{eqn-62}, where $T_{1}=T_{2}=0.015$ and $\Delta_{1}=\Delta_{2}=0.0025$.}
\label{fig-4}
\end{figure}

Here we consider the case that the communication between the controller and the plant is via the ETMs and two communication networks. The controller is applied via the ZOH devices and the networks are assumed to have $\ell_{i}=3$ nodes for $\mathfrak{q}^{i1}_{\p}$, $\mathfrak{q}^{i2}_{\p}$ and $u_{i}$, respectively. In this case, the applied feedback controller is given by $u^{i}_{\ct}=
-c^{-1}_{i}[a_{i}(\sin(\hat{\mathfrak{q}}^{i1}_{\p})-\sin(\hat{\mathfrak{q}}^{i1}_{\rf}))-(\hat{\mathfrak{q}}^{i1}_{\p}-\hat{\mathfrak{q}}^{i1}_{\rf})+(\hat{\mathfrak{q}}^{i2}_{\p}-\hat{\mathfrak{q}}^{i2}_{\rf})]$. $u^{i}_{\f}$ is assumed to be transmitted to \eqref{eqn-61} directly, and $\hat{\mathfrak{q}}^{i1}_{\rf}, \hat{\mathfrak{q}}^{i2}_{\rf}$ are implemented in the ZOH fashion. Hence, $u^{i}_{\ct}$ knows but does not depend on $\mathfrak{q}^{i1}_{\rf}, \mathfrak{q}^{i2}_{\rf}$. In addition, we consider both the RR and TOD protocol cases. In the RR protocol case, the node order of the network 1 is $1\rightarrow2\rightarrow3$ and the node order of the network 2 is $3\rightarrow2\rightarrow1$. In the TOD protocol case, the node to access to each network is the one with a minimum index and the largest norm of the local network-induced error; see Section \ref{subsec-protocol}.

Based on \eqref{eqn-60}-\eqref{eqn-61}, we obtain that $F_{\eta}=(F^{1}_{\eta}, F^{2}_{\eta})$ with $F^{i}_{\eta}=(\eta_{i2}, -a_{i}[\sin(\eta_{i1}+\mathfrak{q}^{i1}_{\rf})-\sin(\mathfrak{q}^{i1}_{\rf})-\sin(\eta_{i1}
+\mathfrak{q}^{i1}_{\rf}+e^{i1}_{\eta}+e^{i1}_{\rf})+\sin(\mathfrak{q}^{i1}_{\rf}+e^{i1}_{\rf})]-(\eta_{i1}+e^{i1}_{\eta})-(\eta_{i2}+e^{i2}_{\eta})+\sum_{j=1, 2}b_{ij}(\eta_{1j}-\eta_{2j})+c_{i}e^{i}_{\f}+c_{i}e^{i}_{\ct})$, $F_{\rf}=(F^{1}_{\rf}, F^{2}_{\rf})$ with $F^{i}_{\rf}=(\mathfrak{q}^{i2}_{\rf}, -a_{i}\sin\mathfrak{q}^{i1}_{\rf}+\sum_{j=1, 2}b_{ij}(\mathfrak{q}^{1j}_{\rf}-\mathfrak{q}^{2j}_{\rf})+c_{i}u^{i}_{\f})$, $G_{\ax}=(-F_{\eta}, 0)$, $G_{\rf}=-F_{\rf}$ and $G_{\f}=-(\dot{u}^{1}_{\f}, \dot{u}^{2}_{\f})$. In addition, $|G^{i}_{\ax}|\leq D_{i}|e_{i}|+|\eta_{i2}|+|(b_{i1}-1)\eta_{i1}+(b_{i2}-1)\eta_{i2}|+|b_{i1}\eta_{(3-i)1}|+|b_{i2}\eta_{(3-i)2}|+2a_{i}|e^{i}_{\rf}|+c_{i}|e^{i}_{\f}|$ with $D_{i}=\sqrt{3}\max\{1+a_{i}, c_{i}\}$. From \cite{Ren2019tracking}, we choose the appropriate Lyapunov function $W_{i}(e_{i}, m_{i}, \kappa_{i}, \tau_{i}, b_{i})$. For instance, $W_{i}(e_{i}, m_{i}, \kappa_{i}, \tau_{i}, b_{i}):=|e^{i}_{\ax}|$ for the TOD protocol. $|\partial W(e_{i}, m_{i}, \kappa_{i}, \tau_{i}, b_{i})/\partial e_{i}|\leq M_{i}$ with $M_{i}=\sqrt{\ell_{i}}$ for the RR protocol case and $M_{i}=1$ for the TOD protocol case. Thus, Assumption \ref{asn-4} holds with $\lambda_{i}=\sqrt{(\ell_{i}-1)/\ell_{i}}$ and $\alpha_{3i}=\alpha_{4i}=\alpha_{5i}=\alpha_{6i}=0$. Assumption \ref{asn-5} holds with $L_{i0}=M_{i}D_{i}$, $L_{i1}=M^{2}_{i}D_{i}/\lambda_{i}$, $H_{i0}(x, e)=H_{i1}(x, e)=M_{i}(|\eta_{i2}|+|(b_{i1}-1)\eta_{i1}+(b_{i2}-1)\eta_{i2}|+|b_{i1}\eta_{(3-i)1}|+|b_{i2}\eta_{(3-i)2}|)$, $\sigma_{1i0}(v)=\sigma_{1i1}(v)=c_{i}M_{i}v$, $\sigma_{2i0}(v)=\sigma_{2i1}(v)=2a_{i}M_{i}v$, and $\sigma_{3i0}(v)=\sigma_{3i1}(v)=0$ for $v\geq0$.

\begin{figure}[!t]
\begin{center}
\begin{picture}(80, 115)
\put(-80, -25){\resizebox{80mm}{50mm}{\includegraphics[width=2.5in]{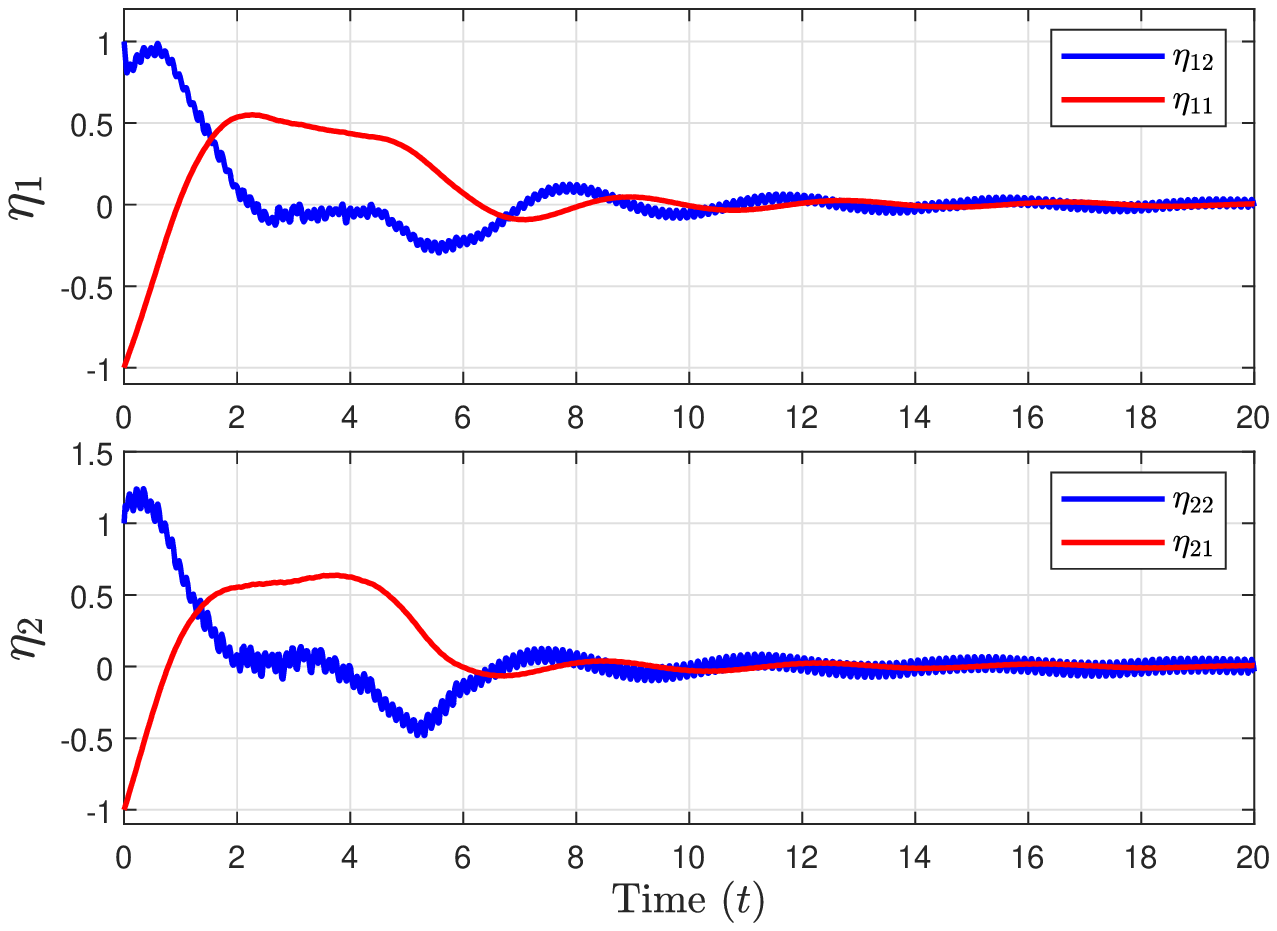}}}
\end{picture}
\end{center}
\caption{Tracking errors under the TOD protocol case and the ETMs \eqref{eqn-62}, where $T_{1}=T_{2}=0.015$ and $\Delta_{1}=\Delta_{2}=0.003$.}
\label{fig-5}
\end{figure}

To verify Assumption \ref{asn-6}, define $V(\eta):=\sum^{2}_{i=1}\phi_{i1}\eta^{2}_{i1}+\phi_{i2}\eta_{i1}\eta_{i2}+\phi_{i3}\eta^{2}_{i2}$, where $\phi_{i1}, \phi_{i2}, \phi_{i3}$ are chosen to make $V$ satisfy \eqref{eqn-29}. Assume that there exist time-varying parameters $\hat{a}_{i}, \tilde{a}_{i}\in[-a_{i}, a_{i}]$ such that $a_{i}[\sin(\eta_{i1}+\mathfrak{q}^{i1}_{\rf})-\sin(\eta_{i1}+\mathfrak{q}^{i1}_{\rf}+e^{i1}_{\eta}+e^{i1}_{\rf})]=\hat{a}_{i}(e^{i1}_{\eta}+e^{i1}_{\rf})$ and
$a_{i}[\sin(\mathfrak{q}^{i1}_{\rf})-\sin(\mathfrak{q}^{i1}_{\rf}+e^{i1}_{\rf})]=\tilde{a}_{i}e^{i1}_{\rf}$. Thus, using twice the fact that $2xy\leq cx^{2}+y^{2}/c$ for all $x, y\geq0$ and $c>0$, we get that $\langle\nabla V(\eta), F_{\eta}(\delta, x, e, w)\rangle\leq\sum^{2}_{i=1}[-\phi_{i1}\eta^{2}_{i1}+(2\phi_{i1}-2\phi_{i3}-\phi_{i2})\eta_{i1}\eta_{i2}-(2\phi_{i3}-\phi_{i1})\eta^{2}_{i2}+
(\phi_{i2}\eta_{i1}+2\phi_{i3}\eta_{i2})(b_{i1}(\eta_{11}-\eta_{21})+b_{i2}(\eta_{12}-\eta_{22}))+0.5(\varrho^{-1}_{i0}+\varrho^{-1}_{i1})(\phi_{i1}\eta_{i1}+2\phi_{i3}\eta_{i2})^{2}
+0.5\varrho_{i0}D_{i}|e_{i}|^2+\varrho_{i1}(4a^{2}_{i}|e^{i}_{\rf}|^{2}+c^{2}_{i}|e^{i}_{\f}|^{2})]$, where $\varrho_{i0}, \varrho_{i1}>0$ are defined in \eqref{eqn-35}. Therefore, if $\phi_{1}, \phi_{2}, \phi_{3}$ are chosen such that \eqref{eqn-29} holds and $-H^{2}_{ib_{i}}(x, e)-K_{ib_{i}}(x, e, m, w)-\varphi_{ib_{i}}(z_{i})\geq-\phi_{i1}\eta^{2}_{i1}+(2\phi_{i1}-2\phi_{i3}-\phi_{i2})\eta_{i1}\eta_{i2}
-(2\phi_{i3}-\phi_{i1})\eta^{2}_{i2}+(\phi_{i1}\eta_{i1}+2\phi_{i3}\eta_{i2})(b_{i1}(\eta_{11}-\eta_{21})+b_{i2}(\eta_{12}-\eta_{22}))+0.5(\varrho^{-1}_{i0}+\varrho^{-1}_{i1})(\phi_{i1}\eta_{i1}
+2\phi_{i3}\eta_{i2})^{2}$, then Assumption \ref{asn-6} is verified with $\theta_{ib_{i}}(v)=\pi_{i}v^{2}$, $\gamma_{i0}=\sqrt{\pi_{i}+\varrho_{i0}D^{2}_{i}}$, $\gamma_{i1}=\sqrt{\pi_{i}+\varrho_{i1}\ell_{i}D^{2}_{i}/\lambda^{2}_{i}}$, $\zeta_{1ib_{i}}(v)=\varrho_{i1}a^{2}_{i}|v|^{2}$, $\zeta_{4ib_{i}}(v)=\varrho_{i1}a^{2}_{i}|v|^{2}$ and $\pi_{i}>0$ is arbitrarily small.

To satisfy the aforementioned conditions, we choose $\phi_{11}=8, \phi_{12}=12, \phi_{13}=6, \phi_{21}=5, \phi_{22}=7, \phi_{23}=9$, $a_{1}=9.81*0.2$, $a_{2}=9.81*0.3$, $c_{1}=2$, $c_{2}=4$, $\pi_{i}=0.005$, $\varrho_{i0}=20$ and $\varrho_{i1}=\varrho_{i0}M_{i}/\lambda_{1}$. Thus, $L_{10}=8.8860$, $L_{11}=18.8501$, $L_{20}=12$, $L_{21}=25.4558$, $\gamma_{10}=22.9436$, $\gamma_{11}=53.8629$, $\gamma_{20}=30.9839$, $\gamma_{21}=72.7386$ for the RR protocol; and $L_{10}=5.1303$, $L_{11}=10.8831$, $L_{20}=6.9282$, $L_{21}=14.6969$, $\gamma_{10}=22.9436$, $\gamma_{11}=31.0978$, $\gamma_{20}=30.9839$, $\gamma_{21}=41.9956$ for the TOD protocol. By the detailed computation, we have that $\overbar{\rho}_{1}=0.0501$ and $\overbar{\rho}_{2}=0.0371$ for RR and TOD protocols. Hence, $\rho_{i}\in(0,\overbar{\rho}_{i})$, and the event-triggered conditions are
\begin{align}
\label{eqn-62}
\Gamma_{i}(\eta_{i}, e_{i})&=\gamma_{i}|(e^{i}_{\eta}, e^{i}_{\rf})|^{2}-\rho_{i}\bar{\lambda}_{i}|\eta_{i}|^{2}\geq0.
\end{align}

\begin{figure}[!t]
\begin{center}
\begin{picture}(80, 115)
\put(-80, -25){\resizebox{80mm}{50mm}{\includegraphics[width=2.5in]{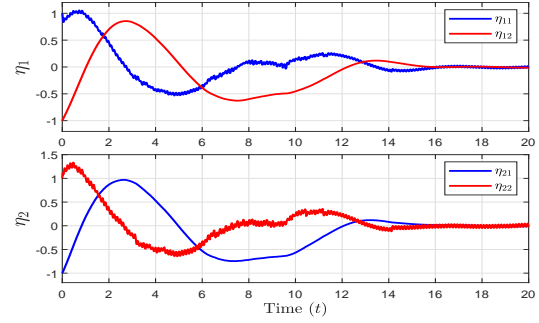}}}
\end{picture}
\end{center}
\caption{Tracking errors under the mixed RR and TOD protocol case and the ETMs \eqref{eqn-62}, where $T_{1}=T_{2}=0.015$ and $\Delta_{1}=\Delta_{2}=0.003$.}
\label{fig-6}
\end{figure}

Set $\phi_{10}(0)=\phi_{11}(0)=1.0956$ and $\phi_{20}(0)=\phi_{21}(0)=0.8774$ for the RR protocol case, and we have $T_{1}=0.0252$, $\Delta_{1}=0.0064$, $T_{2}=0.01605$, and $\Delta_{2}=0.0029$. Set $\phi_{10}(0)=\phi_{11}(0)=\phi_{20}(0)=\phi_{21}(0)=1.0419$ for the TOD protocol case, and we have $T_{1}=0.02795$, $\Delta_{1}=0.0067$, $T_{2}=0.0211$, and $\Delta_{2}=0.0049$. To simplify the simulation, the transmission intervals and the transmission delays are constants. Under the designed event-triggered condition \eqref{eqn-62}, Figs. \ref{fig-4}-\ref{fig-5} show the convergence and boundedness of the tracking errors for both RR and TOD protocol cases, respectively. In addition, different networks are allowed to have different time-scheduling protocols. If the time-scheduling protocol in the network 1 is the RR protocol and the time-scheduling protocol in the network 2 is the TOD protocol, then Fig. \ref{fig-6} shows the convergence of the tracking errors in this mixed protocol case.

\subsection{Robust Distributed Estimation}
\label{subsec-example2}

Consider the following linear plant, which is borrowed from \cite{Li2017robust} and can be treated as the leader agent,
\begin{align}
\label{eqn-63}
\dot{x}_{\p}=Ax_{\p}+Bw,
\end{align}
where $x_{\p}=(x^{1}_{\p}, x^{2}_{\p}, x^{3}_{\p})\in\mathbb{R}^{3}$ is the plant state, and $w\in\mathbb{R}^{3}$ is the external disturbance. Assume that the plant \eqref{eqn-63} has the oscillatory dynamics for $(x^{1}_{\p}, x^{2}_{\p})\in\mathbb{R}^{2}$ and the trivial dynamics for $x^{3}_{\p}\in\mathbb{R}$. Therefore, we take $A=\begin{bmatrix}\begin{smallmatrix}0 &-1 &0 \\ 1 & 0& 0 \\ 0& 0& 0\end{smallmatrix}\end{bmatrix}$ and $B=\diag\{1, 0.5, -1\}$. The outputs of \eqref{eqn-63} are $y=(y^{1}_{\p}, y^{2}_{\p}):=(C_{1}x_{\p}, C_{2}x_{\p})$ with $C_{1}=(1, 1, 0)$ and $C_{2}=(0, 0, 1)$.

According the outputs of the plant \eqref{eqn-63}, we design two observers, which can be treated as the follower agents and assumed to be all-to-all connected, i.e., $\mathcal{G}_{\ob}=\begin{bmatrix}\begin{smallmatrix} 0 & 1 \\ 1& 0\end{smallmatrix}\end{bmatrix}$. These two observers are design as the following form ($i=1, 2$)
\begin{align}
\label{eqn-64}
\dot{x}^{i}_{\ob}&=Ax^{i}_{\ob}+\vartheta_{i}, \quad \dot{\vartheta}_{i}=0, \quad  y^{i}_{\ob}=C_{i}x^{i}_{\ob},
\end{align}
with the update of $\vartheta_{i}\in\mathbb{R}^{3}$ given by ($k\in\mathcal{N}_{i}$)
\begin{align}
\label{eqn-65}
\vartheta^{+}_{i}&=J_{i}(y^{i}_{\ob}-y^{i}_{\p})+J_{ik}(y^{k}_{\ob}-y^{k}_{\p})+\chi_{i}(x^{i}_{\ob}-x^{k}_{\ob}).
\end{align}
Since $(C_{1}, A)$ and $(C_{2}, A)$ are not detectable, each follower agent cannot estimate the full state of the plant by using an observer without using the information from the other follower agent. Hence, two observers in \eqref{eqn-64} are allowed to communicate with each other to reconstruct the state $x_{\p}$.

\begin{figure}[!t]
\begin{center}
\begin{picture}(60, 100)
\put(-55, -18){\resizebox{60mm}{40mm}{\includegraphics[width=2.5in]{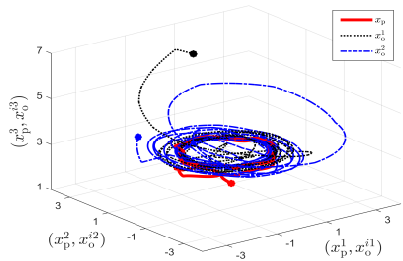}}}
\end{picture}
\end{center}
\caption{The state trajectories of \eqref{eqn-63}-\eqref{eqn-64} under the designed ETM \eqref{eqn-66}. Here, $x_{\p}=(x^{1}_{\p}, x^{2}_{\p}, x^{3}_{\p})$ and $x^{i}_{\ob}=(x^{i1}_{\ob}, x^{i2}_{\ob}, x^{i3}_{\ob})$, for $i \in\{1, 2\}$.}
\label{fig-7}
\end{figure}

\begin{table}[!t]
\centering
\caption[caption]{MASPs for two networks under different $\rho_{i}/\bar{\rho}_{i}$}
\label{tab-1}
\begin{tabular}{c|c|c|c|c|c|c}
\hline \xrowht{6pt}
$\rho_{i}/\bar{\rho}_{i}$ & 0& 0.1 & 0.2 & 0.3 & 0.4 & 0.5  \\  \hline
$T_{1}$ & 0.5344 & 0.4781  & 0.4208 & 0.3634 & 0.3067 & 0.2511 \\ \hline
$T_{2}$ & 0.7550 & 0.6418  & 0.5416 & 0.4518 & 0.3704 & 0.2962 \\ \hline
$\rho_{i}/\bar{\rho}_{i}$  & 0.6 & 0.7 & 0.8 & 0.9 & 1   \\  \hline
$T_{1}$  & 0.1971 & 0.1448 & 0.0945 & 0.0462 & 0 \\ \hline
$T_{2}$  & 0.2279 & 0.1648 & 0.1061 & 0.0514 & 0 \\ \hline
\end{tabular}
\end{table}

To compare with \cite{Li2017robust}, we choose the parameters in \eqref{eqn-65} as: $J_{1}=(-0.5, -0.2, -0.1)$, $J_{12}=(-0.2, -0.2, -0.5)$, $J_{21}=(0.2, 0.3, 0.3)$, $J_{2}=(-0.1, -0.5, 0.2)$, and $\chi_{1}=\chi_{2}=-0.4$. Define $\eta_{i}:=x^{i}_{\ob}-x_{\p}$, $e^{i}_{\p}:=\hat{y}^{i}_{\p}-y^{i}_{\p}$, $e^{i}_{\ob}:=\hat{y}^{i}_{\ob}-y^{i}_{\ob}$, $\psi_{i}:=J_{i}(y^{i}_{\ob}-y^{i}_{\p})+J_{ik}(y^{k}_{\ob}-y^{k}_{\p})+\chi_{i}(x^{i}_{\ob}-x^{k}_{\ob})-\vartheta_{i}$, and $e^{i}_{\ax}:=e^{i}_{\ob}-e^{i}_{\p}$. Note that $y_{1}, y_{2}\in\mathbb{R}$, and thus $e^{i}_{\p}=e^{i}_{\ob}=e^{i}_{\ax}=0$ for the successful transmission. Hence, we define the functions $V_{i}:=\eta^{\tran}_{i}\eta_{i}$, $W_{i}:=|\psi_{i}|$ and $\varphi(y_{i})=0.5y^{2}_{i}$ for $i=1, 2$. By the detailed computation, Assumption \ref{asn-4} is verified with $\lambda_{1}=\lambda_{2}=0$; Assumption \ref{asn-5} is verified with $L_{1}=1.8142$, $L_{2}=1.4$, $H_{1}(x, e):=|C_{1}||(A+J_{1}C_{1}+\chi_{1}I)\eta_{1}+(J_{12}C_{2}-\chi_{1}I)\eta_{2}+J_{12}e^{2}_{\ob}|+|\chi_{1}||(A+(J_{1}-J_{21})C_{1}
+(\chi_{1}+\chi_{2})I)\eta_{1}-(A+(J_{2}-J_{12})C_{2}+(\chi_{1}+\chi_{2})I)\eta_{2}+(J_{12}-J_{2})e^{2}_{\ob}-\psi_{2}|$, and $H_{2}(x, e):=|C_{2}||(A+J_{2}C_{2}+\chi_{2}I)\eta_{2}
+(J_{21}C_{1}+\chi_{2}I)\eta_{1}+J_{21}e^{1}_{\ob}|+|\chi_{2}||(A+(J_{2}-J_{12})C_{2}+(\chi_{1}+\chi_{2})I)\eta_{2}-(A+(J_{1}-J_{21})C_{1}+(\chi_{1}+\chi_{2})I)\eta_{1}
+(J_{21}-J_{1})e^{1}_{\ob}-\psi_{1}|$. Assumption \ref{asn-8} holds with $\mu_{1}=\theta_{1}=-0.4$, $\mu_{2}=\theta_{2}=-0.2$, $\gamma_{1}=1.7243$, $\gamma_{2}=1.5045$, $\bar{L}_{1}=\bar{L}_{2}=0$, $K_{1}(x, e):=|\eta_{1}C^{\tran}_{1}C_{1}[(J_{12}C_{2}-\chi_{1}I)\eta_{2}+J_{1}e^{1}_{\ob}+J_{12}e^{2}_{\ob}-\psi_{1}]|$, and $K_{2}(x, e):=|\eta_{2}C^{\tran}_{2}C_{2}[(J_{21}C_{1}-\chi_{2}I)\eta_{1}+J_{2}e^{2}_{\ob}+J_{21}e^{1}_{\ob}-\psi_{2}]|$. Based on aforementioned values, we have that $\overbar{\rho}_{1}=0.5799$ and $\overbar{\rho}_{2}=0.6647$. In addition, for $i=1, 2$, $\overbar{\lambda}_{i}\in[0, 1)$, $\rho_{i}\in(0,\overbar{\rho}_{i})$, and the event-triggered conditions are given by
\begin{align}
\label{eqn-66}
\Gamma_{i}(\eta_{i}, e^{i}_{\ax}, \psi_{i})&=\gamma_{i}|\psi_{i}|^{2}-\rho_{i}\bar{\lambda}_{i}|C_{i}\eta_{i}|^{2}\geq0.
\end{align}
Obviously, different choices of $\overbar{\lambda}_{i}$ and $\rho_{i}$ lead to different event-triggered conditions. Under different values of $\rho_{i}$, the MASP $T_{i}$ is computed and illustrated in Table \ref{tab-1}.

\begin{figure}[!t]
\begin{center}
\begin{picture}(70,100)
\put(-70,-15){\resizebox{70mm}{40mm}{\includegraphics[width=2.5in]{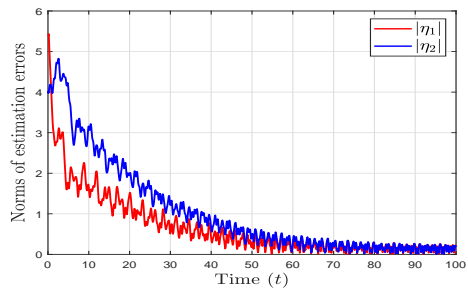}}}
\end{picture}
\end{center}
\caption{The evolution of the norms of the estimation errors.}
\label{fig-8}
\end{figure}

From Table \ref{tab-1}, we choose $\rho_{i}=0.2\overbar{\rho}_{i}$ and set the MASPs $T_{1}=0.2$ and $T_{2}=0.4$. Given the initial states $x_{\p}(0)=(1, 1, 1)$, $x^{1}_{\ob}(0)=(1, 3, 6)$, $x^{2}_{\ob}(0)=(-2, 2, 3.5)$, $\vartheta_{1}(0)=(1, 1, 1)$, and $\vartheta_{2}(0)= (-1, -1, -1)$. In the PETC case, the state trajectories of the plant and the observers are shown in Fig. \ref{fig-7}, which implies the convergence of the estimates $x^{1}_{\ob}=(x^{11}_{\ob}, x^{12}_{\ob}, x^{13}_{\ob})$ and $x^{2}_{\ob}=(x^{21}_{\ob}, x^{22}_{\ob}, x^{23}_{\ob})$ to $x_{\p}=(x^{1}_{\p}, x^{2}_{\p}, x^{3}_{\p})$. Fig. \ref{fig-8} shows the evolution of the norms of two estimation errors, and implies the convergence of the estimation errors.

Comparing with \cite{Li2017robust, Park2016design} on time-triggered observers, the distributed event-triggered observers are considered in our setting. Note that the discrete-time LTI system is addressed in \cite{Park2016design} and that the sampling periods are the same for all observers. On the other hand, even though the data in this example would satisfy the conditions in \cite{Li2017robust}, the MASPs here are computed instead of given \emph{a priori}. In addition, due to the designed event-triggered condition \eqref{eqn-66}, the numbers of the event-triggering times are reduced, that is, 392 times for the first observer and 204 times for the second observer (in 100 units of time), whereas the corresponding event-triggering numbers are 500 and 250 in \cite{Li2017robust}.

\section{Conclusions}
\label{sec-conclusion}

We presented a Lyapunov-based emulation approach for the event-triggered tracking control problem of networked MAS, where external disturbances are considered and the information communication is via multiple asynchronous networks. To deal with the considered problem, we proposed a new hybrid model, and then established sufficient conditions and designed decentralized event-triggered mechanisms. The tradeoff between the MASP and the MAD was determined to guarantee the tracking performance. In addition, we showed the direct employment of these obtained results to derive new results for the event-triggered observer design problem for networked MAS. The effectiveness of the proposed approach was illustrated via two numerical examples.

Many extensions of the obtained results can be envisioned in different directions. The results can be extended to the case of LTI systems for the co-design of the event-triggering mechanisms and the decentralized controllers/observers. The event-triggered tracking control for MAS under limited data rate can be studied by modifying the derived model appropriately and then combining the techniques in \cite{Ren2019tracking} and this paper.

\appendices
\section{Explicit Expression in \eqref{eqn-17a}}
\label{sec-appendixA}

The functions in \eqref{eqn-17a} are presented below in detail.
\begin{align*}
\dot{\eta}&=f_{\p}(x_{\p}, \hat{u}_{\ct}+\hat{u}_{\f}, w_{\p})-f_{\p}(x_{\rf}, \hat{u}_{\f}, w_{\rf})\\
&=f_{\p}(\eta+x_{\rf}, g_{\ct}(x_{\ct})+e_{\ct}+u_{\f}+e_{\f}, w_{\p}) \\
&\quad -f_{\p}(x_{\rf}, u_{\f}+e_{\f}, w_{\rf})\\
&=:F_{\eta}(\delta, \eta, x_{\rf}, x_{\ct}, e_{\ax}, e_{\rf}, e_{\f}, w),
\end{align*}
where $\delta$ is a variable related to $u_{\f}$ and/or it differential.
\begin{align*}
\dot{x}_{\rf}&=f_{\p}(x_{\rf}, \hat{u}_{\f}, w_{\rf})=f_{\p}(x_{\rf}, u_{\f}+e_{\f}, w_{\rf}) \\
&=:F_{\rf}(\delta, \eta, x_{\rf}, x_{\ct}, e_{\ax}, e_{\rf}, e_{\f}, w), \\
\dot{x}_{\ct}&=f_{\ct}(x_{\ct}, \hat{y}_{\p}, \hat{y}_{\rf}, w_{\ct}) \\
&=f_{\ct}(x_{\ct}, g_{\p}(\eta+x_{\rf})+e_{\p}, g_{\rf}(x_{\rf})+e_{\rf}, w_{\ct}) \\
&=:F_{\ct}(\delta, \eta, x_{\rf}, x_{\ct}, e_{\ax}, e_{\rf}, e_{\f}, w), \\
\dot{e}_{\eta}&=\dot{e}_{\p}-\dot{e}_{\rf} \\
&=-\langle\nabla g_{\p}(x_{\p}), f_{\p}(\eta+x_{\rf}, g_{\ct}(x_{\ct})+e_{\ct}+u_{\f}+e_{\f}, w_{\p})\rangle \\
&\quad +\langle\nabla g_{\p}(x_{\rf}), f_{\p}(x_{\rf}, \hat{u}_{\ct}+\hat{u}_{\f}, w_{\rf})\rangle, \\
\dot{e}_{\ct}&=-\langle\nabla g_{\ct}(x_{\ct}), f_{\ct}(x_{\ct}, g_{\p}(\eta+x_{\rf})+e_{\p}, g_{\rf}(x_{\rf})+e_{\rf}, w_{\ct})\rangle,\\
\dot{e}_{\ax}&=\begin{bmatrix} \dot{e}_{\eta}\\ \dot{e}_{\ct} \end{bmatrix}
=\begin{bmatrix} \dot{e}_{\p}-\dot{e}_{\rf} \\ \dot{e}_{\ct}\end{bmatrix}=:G_{\ax}(\delta, \eta, x_{\rf}, x_{\ct}, e_{\ax}, e_{\rf}, e_{\f}, w), \\
\dot{e}_{\rf}&=-\langle\nabla g_{\p}(x_{\rf}), f_{\p}(x_{\rf}, u_{\f}+e_{\f}, w_{\rf})\rangle \\
&=:G_{\rf}(\delta, \eta, x_{\rf}, x_{\ct}, e_{\ax}, e_{\rf}, e_{\f}, w), \\
\dot{e}_{\f}&=-\dot{u}_{\f}=:G_{\f}(\delta, \eta, x_{\rf}, x_{\ct}, e_{\ax}, e_{\rf}, e_{\f}, w).
\end{align*}

\setcounter{equation}{0}
\renewcommand{\theequation}{B.\arabic{equation}}

\section{Proofs of Main Results}
\label{sec-appendixB}

\subsection{Proof of Theorem \ref{thm-1}}

For $\mathfrak{X}\in C\cup D\cup G(D)$, define the function as
\begin{align}
\label{eqn-B1}
U(\mathfrak{X})&:=V(x)+\sum^{N}_{i=1}\mathcal{W}_{i}(\mathfrak{X}),
\end{align}
with $\mathcal{W}_{i}(\mathfrak{X}):=\max\{\gamma_{ib_{i}}\phi_{ib_{i}}(\tau_{i})W^{2}_{i}(e_{i}, m_{i}, \kappa_{i}, b_{i}), (1-b_{i})\rho_{i}\varphi_{ib_{i}}(z_{i})\}$ and $i\in\mathcal{N}$, where $\gamma_{ib_{i}}, W_{i}$ and $\varphi_{ib_{i}}$ are given in Assumptions \ref{asn-3}-\ref{asn-4}, and $\rho_{i}$ is given in \eqref{eqn-32}. In the following, we first show that $U(\mathfrak{X})$ is suitable Lyapunov function for the system \eqref{eqn-18} (i.e., Steps 1-3), and then derive the convergence of $U(\mathfrak{X})$ along the hybrid time line (i.e., Step 4).

\textbf{Step 1: Positive Definiteness and Radial Unboundedness of $U(\mathfrak{X})$.}
If $b_{i}=1$, then $\mathcal{W}_{i}(\mathfrak{X})=\gamma_{i1}\phi_{i1}(\tau_{i})W^{2}_{i}(e_{i}, m_{i}, \kappa_{i}, 1)$ and thus from \eqref{eqn-24} and \eqref{eqn-35}, one has
\begin{align}
\label{eqn-B2}
\mathcal{W}_{i}(\mathfrak{X})&\geq\gamma_{i1}\bar{\lambda}_{i}\alpha^{2}_{1i}(|e^{i}_{\ax}|).
\end{align}
If $b_{i}=0$, then $\mathcal{W}_{i}(\mathfrak{X}):=\max\{\gamma_{i0}\phi_{i0}(\tau_{i})W^{2}_{i}(e_{i}, m_{i}, \kappa_{i}, 0),$ $\rho_{i}\varphi_{i0}(z_{i})\}$.
From Assumption \ref{asn-6}, $\varphi_{ib_{i}}$ is locally Lipschitz and positive definite and $\varphi_{ib_{i}}(0)=0$. In addition, $z_{i}=(y^{i}_{\p}, y^{i}_{\rf}, u^{i}_{\f}, u^{i}_{\ct})=(g^{i}_{\p}(x^{i}_{\p}), g^{i}_{\p}(x^{i}_{\rf}), u^{i}_{\f}, g^{i}_{\ct}(x^{i}_{\ct}))$, where $g^{i}_{\p}, g^{i}_{\p}$ and $g^{i}_{\ct}$ are continuous differential. Hence, there exits $\bar{\alpha}_{i}\in\mathcal{K}$ such that $\varphi_{ib_{i}}(z_{i})\leq\tilde{\alpha}_{i}(|(x, e)|)$, and from \eqref{eqn-24} and \eqref{eqn-36},
\begin{align}
\label{eqn-B3}
\mathcal{W}_{i}(\mathfrak{X})&\leq\bar{\lambda}^{-1}_{i}\gamma_{i0}\alpha^{2}_{2i}(|e_{i}|)+\rho_{i}\tilde{\alpha}_{i}(|(x, e)|).
\end{align}
From \eqref{eqn-B2}-\eqref{eqn-B3} and \eqref{eqn-29}, there exist $\alpha_{1}, \alpha_{2}\in\mathcal{K}_{\infty}$ such that
\begin{align}
\label{eqn-B4}
\alpha_{1}(|(\eta, e_{\ax})|)\leq U(\mathfrak{X})\leq\alpha_{2}(|\mathfrak{X}|),
\end{align}
where $\alpha_{1}(v):=\alpha_{1V}(v/2)+\sum^{N}_{i=1}\gamma_{i1}\bar{\lambda}_{i}\alpha^{2}_{1i}(v/2)$ and
$\alpha_{2}(v):=\alpha_{2V}(v)+\sum^{N}_{i=1}[\bar{\lambda}^{-1}_{i}\gamma_{i0}\alpha^{2}_{2i}(v)+\rho_{i}\tilde{\alpha}_{i}(v)]$.

\textbf{Step 2: Decreasing of $U(\mathfrak{X})$ on the Flow.} From the definition of $U(\mathfrak{X})$, we consider the following two cases.

\subsubsection*{Case 1} $\gamma_{ib_{i}}\phi_{ib_{i}}(\tau_{i})W^{2}_{i}(e_{i}, m_{i}, \kappa_{i}, b_{i})\geq\rho_{i}\varphi_{ib_{i}}(z_{i})$. For the flow equation $F$ in \eqref{eqn-20}, we have\footnote{$\langle\nabla U(\mathfrak{X}), F(\mathfrak{X}, w)\rangle$ is used with a slight abuse of terminology since $U$ is not differential almost everywhere. This is justified by  $\dot{\kappa}_{i}=\dot{b}_{i}=0$ in \eqref{eqn-20}.}
\begin{align}
\label{eqn-B5}
&\langle\nabla U(\mathfrak{X}), F(\mathfrak{X}, w)\rangle=\langle\nabla V(x), f(\delta, x, e, w)\rangle \nonumber\\
&\quad+\sum^{N}_{i=1}[\gamma_{ib_{i}}\dot{\phi}_{ib_{i}}(\tau_{i})W^{2}_{i}(e_{i}, m_{i}, \kappa_{i}, b_{i}) \nonumber \\
&\quad+2\gamma_{ib_{i}}\phi_{ib_{i}}(\tau_{i})W_{i}(e_{i}, m_{i}, \kappa_{i}, b_{i}) \nonumber \\
&\quad\left.\times\left\langle\frac{\partial W_{i}(e_{i}, m_{i}, \kappa_{i}, b_{i})}{\partial e_{i}}, g_{i}(\delta, x, e, w)\right\rangle\right] \nonumber \\
&\leq-\mu V(x)+\sum^{N}_{i=1}[\Pi_{i}(\mathfrak{X})-\theta_{ib_{i}}W^{2}_{i}(e_{i}, m_{i}, \kappa_{i}, b_{i})],
\end{align}
where
\begin{align}
\label{eqn-B6}
\Pi_{i}(\mathfrak{X})&:=\gamma^{2}_{ib_{i}}W^{2}_{i}(e_{i}, m_{i}, \kappa_{i}, b_{i})-H^{2}_{ib_{i}}(x, e)-K_{i}(x, e, m, w) \nonumber\\
&\quad-\varphi_{i}(z_{i})+\zeta_{1ib_{i}}(|e^{i}_{\f}|)+\zeta_{2ib_{i}}(|e^{i}_{\rf}|)+\zeta_{3ib_{i}}(|w|)\nonumber\\
&\quad+\gamma_{ib_{i}}[-2L_{ib_{i}}\phi_{ib_{i}}(\tau_{i})-\gamma_{ib_{i}}((1+\varrho_{ib_{i}})\phi^{2}_{ib_{i}}(\tau_{i})+1)] \nonumber\\
&\quad\times W^{2}_{i}(e_{i}, m_{i}, \kappa_{i}, b_{i})+2\gamma_{ib_{i}}\phi_{ib_{i}}(\tau_{i}) \nonumber\\
&\quad \times W_{i}(e_{i}, m_{i}, \kappa_{i}, b_{i})[L_{ib_{i}}W_{i}(e_{i}, m_{i}, \kappa_{i}, b_{i}) \nonumber\\
&\quad+H_{ib_{i}}(x, e)+\sigma_{1ib_{i}}(|e^{i}_{\f}|)+\sigma_{2ib_{i}}(|e^{i}_{\rf}|)+\sigma_{3ib_{i}}(|w|)] \nonumber \\
&\leq-H^{2}_{ib_{i}}(x, e)+\zeta_{1ib_{i}}(|e^{i}_{\f}|)+\zeta_{2ib_{i}}(|e^{i}_{\rf}|)+\zeta_{3ib_{i}}(|w|) \nonumber\\
&\quad+\gamma_{ib_{i}}[-2L_{ib_{i}}\phi_{ib_{i}}(\tau_{i})-\gamma_{ib_{i}}(1+\varrho_{ib_{i}})\phi^{2}_{ib_{i}}(\tau_{i})]\nonumber\\
&\quad \times W^{2}_{i}(e_{i}, m_{i}, \kappa_{i}, b_{i})+2\gamma_{ib_{i}}\phi_{ib_{i}}(\tau_{i})W_{i}(e_{i}, m_{i}, \kappa_{i}, b_{i}) \nonumber\\
&\quad\times[L_{ib_{i}}W_{i}(e_{i}, m_{i}, \kappa_{i}, b_{i})+H_{ib_{i}}(x, e)] \nonumber\\
&\quad+\varrho_{ib_{i}}\gamma^{2}_{ib_{i}}\phi^{2}_{ib_{i}}(\tau_{i})W^{2}_{i}(e_{i}, m_{i}, \kappa_{i}, b_{i}) \nonumber\\
&\quad+3\varrho^{-1}_{ib_{i}}[\sigma^{2}_{1ib_{i}}(|e^{i}_{\f}|)+\sigma^{2}_{2ib_{i}}(|e^{i}_{\rf}|)+\sigma^{2}_{3ib_{i}}(|w|)] \nonumber\\
&\leq\zeta_{1ib_{i}}(|e^{i}_{\f}|)+\zeta_{2ib_{i}}(|e^{i}_{\rf}|)+\zeta_{3ib_{i}}(|w|)+3\varrho^{-1}_{ib_{i}} \nonumber\\
&\quad\times[\sigma^{2}_{1ib_{i}}(|e^{i}_{\f}|)+\sigma^{2}_{2ib_{i}}(|e^{i}_{\rf}|)+\sigma^{2}_{3ib_{i}}(|w|)],
\end{align}
where, the first ``$\leq$'' holds because of Assumptions \ref{asn-5}-\ref{asn-6}, and the second ``$\leq$'' holds due to the fact that $2\mathfrak{a}\mathfrak{b}\leq\mathfrak{c}\mathfrak{a}^{2}+\mathfrak{b}^{2}/\mathfrak{c}$ for all $\mathfrak{a}, \mathfrak{b}\geq0, \mathfrak{c}>0$. From \eqref{eqn-B5}-\eqref{eqn-B6}, we have
\begin{align}
\label{eqn-B7}
\langle\nabla U(\mathfrak{X}), F(\mathfrak{X}, w)\rangle&\leq-\mu V(x)-\theta_{1}\sum^{N}_{i=1}W_{i}(e_{i}, m_{i}, \kappa_{i}, b_{i}) \nonumber\\
&\quad +\bar{\sigma}_{1}(|e_{\f}|)+\bar{\sigma}_{2}(|e_{\rf}|)+\bar{\sigma}_{3}(|w|),
\end{align}
where $\theta_{1}:=\min_{i\in\mathcal{N}}\{\gamma^{-1}_{ib_{i}}\bar{\lambda}_{i}\theta_{ib_{i}}\}$ and $\bar{\sigma}_{k}(v):=\sum^{N}_{i=1}\max\{\zeta_{ki0}(v)$ $+3\varrho^{-1}_{i0}\sigma^{2}_{ki0}(v), \zeta_{ki1}(v)+3\varrho^{-1}_{i1}\sigma^{2}_{ki1}(v)\}$ with $k\in\{1, 2, 3\}$.

\subsubsection*{Case 2} $\gamma_{ib_{i}}\phi_{ib_{i}}(\tau_{i})W^{2}_{i}(e_{i}, m_{i}, \kappa_{i}, b_{i})<\rho_{i}\varphi_{ib_{i}}(z_{i})$. In this case, the derivative of $U(\mathfrak{X})$ is given by
\begin{align}
\label{eqn-B8}
&\langle\nabla U(\mathfrak{X}), F(\mathfrak{X}, w)\rangle \nonumber \\
&=\langle\nabla V(x), f(\delta, x, e, w)\rangle+\sum^{N}_{i=1}\rho_{i}\dot{\varphi}_{ib_{i}}(z_{i}) \nonumber \\
&\leq-\mu V(x)+\sum^{N}_{i=1}[\bar{\Pi}_{i}(\mathfrak{X})-\theta_{ib_{i}}W^{2}_{i}(e_{i}, m_{i}, \kappa_{i}, b_{i})],
\end{align}
where
\begin{align}
\label{eqn-B9}
\bar{\Pi}_{i}(\mathfrak{X})&:=\gamma^{2}_{ib_{i}}W^{2}_{i}(e_{i}, m_{i}, \kappa_{i}, b_{i})-H^{2}_{ib_{i}}(x, e)-K_{ib_{i}}(x, e, w) \nonumber\\
&\quad-\varphi_{ib_{i}}(z_{i})+\zeta_{1ib_{i}}(|e^{i}_{\f}|)+\zeta_{2ib_{i}}(|e^{i}_{\rf}|)+\zeta_{3ib_{i}}(|w|)  \nonumber\\
&\quad+\rho_{i}\bar{L}_{ib_{i}}\varphi_{ib_{i}}(z_{i})+\rho_{i}H^{2}_{ib_{i}}(x, e)+\rho_{i}K_{ib_{i}}(x, e, w)  \nonumber \\
&\quad +\rho_{i}\zeta_{4ib_{i}}(|e^{i}_{\f}|)+\rho_{i}\zeta_{5ib_{i}}(|e^{i}_{\rf}|)+\rho_{i}\zeta_{6ib_{i}}(|w|)\nonumber\\
&\leq\gamma^{2}_{ib_{i}}W^{2}_{i}(e_{i}, m_{i}, \kappa_{i}, b_{i})-\varphi_{ib_{i}}(z_{i})+\rho_{i}\bar{L}_{ib_{i}}\varphi_{ib_{i}}(z_{i}) \nonumber\\
&\quad-(1-\rho_{i})[H^{2}_{ib_{i}}(x, e)+K_{ib_{i}}(x, e, w)] \nonumber\\
&\quad +\bar{\zeta}_{1i}(|e_{\f}|)+\bar{\zeta}_{2i}(|e_{\rf}|)+\bar{\zeta}_{3i}(|w|),
\end{align}
where $\bar{\zeta}_{1i}(v):=\max_{b_{i}\in\{0, 1\}}\{\zeta_{1ib_{i}}(v)+\rho_{i}\zeta_{4ib_{i}}(v)\}$, $\bar{\zeta}_{2i}(v):=\max_{b_{i}\in\{0, 1\}}\{\zeta_{2ib_{i}}(v)+\rho_{i}\zeta_{5ib_{i}}(v)\}$ and $\bar{\zeta}_{3i}(v):=\max_{b_{i}\in\{0, 1\}}\{\zeta_{3ib_{i}}(v)+\rho_{i}\zeta_{6ib_{i}}(v)\}$. Since $\gamma_{ib_{i}}\phi_{ib_{i}}(\tau_{i})W^{2}_{i}(e_{i}, m_{i}, \kappa_{i}, b_{i})<\rho_{i}\varphi_{ib_{i}}(z_{i})$, we only need to consider the case $b_{i}=0$, and have
\begin{align}
\label{eqn-B10}
\gamma^{2}_{i0}W^{2}_{i}(e_{i}, m_{i}, \kappa_{i}, 0)&<\phi^{-1}_{i0}(\tau_{i})\gamma_{i0}\rho_{i}\varphi_{i0}(z_{i}) \nonumber \\
&<\bar{\lambda}^{-1}_{i}\gamma_{i0}\rho_{i}\varphi_{i0}(z_{i}) \nonumber \\
&<(1-\rho_{i}\bar{L}_{i0})\varphi_{i0}(z_{i}),
\end{align}
where, the second ``$<$'' holds from \eqref{eqn-36}; the third ``$<$'' holds due to \eqref{eqn-33}-\eqref{eqn-34}. From \eqref{eqn-B9}-\eqref{eqn-B10} and \eqref{eqn-33}-\eqref{eqn-34}, we have $\bar{\Pi}_{i}(\mathfrak{X})\leq\bar{\zeta}_{1}(|e_{\f}|)+\bar{\zeta}_{2}(|e_{\rf}|)+\bar{\zeta}_{3}(|w|)$. Therefore, we have
\begin{align}
\label{eqn-B11}
\langle\nabla U(\mathfrak{X}), F(\mathfrak{X}, w)\rangle&\leq-\mu V(x)-\theta_{2}\sum^{N}_{i=1}W_{i}(e_{i}, m_{i}, \kappa_{i}, b_{i}) \nonumber\\
& +\bar{\zeta}_{1}(|e_{\f}|)+\bar{\zeta}_{2}(|e_{\rf}|)+\bar{\zeta}_{3}(|w|),
\end{align}
where $\theta_{2}:=\min_{i\in\mathcal{N}}\{\theta_{i0}\}$, $\bar{\zeta}_{1}(v):=\sum^{N}_{i=1}\bar{\zeta}_{1i}(v)$, $\bar{\zeta}_{2}(v):=\sum^{N}_{i=1}\bar{\zeta}_{2i}(v)$, and $\bar{\zeta}_{3}(v):=\sum^{N}_{i=1}\bar{\zeta}_{3i}(v)$. Obviously, $\theta_{2}\geq\theta_{1}$.

\textbf{Step 3: Non-increasing of $U(\mathfrak{X})$ at Jumps.} From the definition of $b_{i}$, we consider the following two cases.

\subsubsection*{Case 1} $b_{i}=0$. If the event-triggered condition is satisfied, then $\gamma_{i0}W^{2}_{i}(e_{i}, m_{i}, \kappa_{i}, 0)\geq\rho_{i}\bar{\lambda}_{i}\varphi_{i0}(z_{i})$, combining which with the fact that $\phi_{i0}(\tau_{i})\in[\bar{\lambda}_{i}, \bar{\lambda}^{-1}_{i}]$ (see \cite{Postoyan2014tracking}) yields that $\gamma_{i0}\phi_{i0}(\tau_{i})W^{2}_{i}(e_{i}, m_{i}, \kappa_{i}, 0)\geq\rho_{i}\varphi_{i0}(z_{i})$. Therefore, $\mathcal{W}_{i}(\mathfrak{X}):=\gamma_{i0}\phi_{i0}(\tau_{i})W^{2}_{i}(e_{i}, m_{i}, \kappa_{i}, 0)$ and
\begin{align}
\label{eqn-B13}
&U(G_{1}(\mathfrak{X}))\leq V(x)+\sum^{N}_{i=1}\gamma_{i1}\phi_{i1}(0)[W_{i}(e_{i}, m_{i}, \kappa_{i}, 0) \nonumber\\
&\quad+\alpha_{3i}(|e^{i}_{\f}|)+\alpha_{4i}(|e^{i}_{\rf}|)]^{2}\nonumber\\
&\leq V(x)+\sum^{N}_{i=1}\gamma_{i1}\phi_{i1}(0)[(1+\varrho_{i1})\bar{\lambda}^{2}_{i}W^{2}_{i}(e_{i}, m_{i}, \kappa_{i}, 0)\nonumber\\
&\quad+(1+2\varrho^{-1}_{i1})(\alpha^{2}_{3i}(|e^{i}_{\f}|)+\alpha^{2}_{4i}(|e^{i}_{\rf}|))]\nonumber\\
&\leq V(x)+\sum^{N}_{i=1}\gamma_{i0}\phi_{i0}(\tau_{i})W^{2}_{i}(e_{i}, m_{i}, \kappa_{i}, 0)+\bar{\alpha}_{3}(|e_{\f}|)+\bar{\alpha}_{4}(|e_{\rf}|) \nonumber\\
&=U(\mathfrak{X})+\bar{\alpha}_{3}(|e_{\f}|)+\bar{\alpha}_{4}(|e_{\rf}|),
\end{align}
where, $\bar{\alpha}_{3}(|e_{\f}|):=\sum^{N}_{i=1}(1+2\varrho^{-1}_{i1})\gamma_{i1}\bar{\lambda}^{-1}_{i}\alpha^{2}_{3i}(|e^{i}_{\f}|)$ and $\bar{\alpha}_{4}(|e_{\rf}|):=\sum^{N}_{i=1}(1+2\varrho^{-1}_{i1})\gamma_{i1}\bar{\lambda}^{-1}_{i}\alpha^{2}_{4i}(|e^{i}_{\rf}|)$. If the event-triggered condition is not satisfied, then
\begin{align}
\label{eqn-B14}
U(G_{1}(\mathfrak{X}))&=V(x)+\sum^{N}_{i=1}\rho_{i}\varphi_{i}(z_{i})=U(\mathfrak{X}).
\end{align}

\subsubsection*{Case 2} $b_{i}=1$. In this case, $U(\mathfrak{X})=V(x)+\sum^{N}_{i=1}\gamma_{i1}\phi_{i1}(\tau_{i})W^{2}_{i}(e_{i}, m_{i}, \kappa_{i}, 1)$ and $G(\mathfrak{X})=G_{2}(\mathfrak{X})$.
\begin{align}
\label{eqn-B12}
&U(G_{2}(\mathfrak{X}))\leq V(x)+\sum^{N}_{i=1}\gamma_{i0}\phi_{i0}(\tau_{i})[W_{i}(e_{i}, m_{i}, \kappa_{i}, 1) \nonumber\\
&\quad+\alpha_{5i}(|e^{i}_{\f}|)+\alpha_{6i}(|e^{i}_{\rf}|)]^{2}\nonumber\\
&\leq V(x)+\sum^{N}_{i=1}\gamma_{i0}\phi_{i0}(\tau_{i})[(1+\varrho_{i0})W^{2}_{i}(e_{i}, m_{i}, \kappa_{i}, 1)\nonumber\\
&\quad+(1+2\varrho^{-1}_{i0})(\alpha^{2}_{5i}(|e^{i}_{\f}|)+\alpha^{2}_{6i}(|e^{i}_{\rf}|))]\nonumber\\
&\leq V(x)+\sum^{N}_{i=1}\gamma_{i1}\phi_{i1}(\tau_{i})W^{2}_{i}(e_{i}, m_{i}, \kappa_{i}, 1) \nonumber\\
&\quad +\bar{\alpha}_{5}(|e_{\f}|)+\bar{\alpha}_{6}(|e_{\rf}|) \nonumber\\
&=U(\mathfrak{X})+\bar{\alpha}_{5}(|e_{\f}|)+\bar{\alpha}_{6}(|e_{\rf}|),
\end{align}
where the first ``$\leq$'' holds from Assumption \ref{asn-4}, the second ``$\leq$'' holds from the fact that $2\mathfrak{a}\mathfrak{b}\leq\mathfrak{c}\mathfrak{a}^{2}+\mathfrak{b}^{2}/\mathfrak{c}$ for all $\mathfrak{a}, \mathfrak{b}\geq0, \mathfrak{c}>0$, the third ``$\leq$'' holds from \eqref{eqn-29}, $\bar{\alpha}_{5}(|e_{\f}|):=\sum^{N}_{i=1}(1+2\varrho^{-1}_{i0})\gamma_{i0}\bar{\lambda}^{-1}_{i}\alpha^{2}_{5i}(|e^{i}_{\f}|)$, and $\bar{\alpha}_{6}(|e_{\rf}|):=\sum^{N}_{i=1}(1+2\varrho^{-1}_{i0})\gamma_{i0}\bar{\lambda}^{-1}_{i}\alpha^{2}_{6i}(|e^{i}_{\rf}|)$.

\textbf{Step 4: Convergence along the Hybrid Time Line.} From Steps 2 and 3, we have
\begin{align}
\label{eqn-B15}
\langle\nabla U(\mathfrak{X}), F(\mathfrak{X}, w)\rangle&\leq-\varpi U(\mathfrak{X})+\zeta_{1}(|e_{\f}|) \nonumber\\
&\quad +\zeta_{2}(|e_{\rf}|)+\zeta_{3}(|w|), \\
\label{eqn-B16}
U(\mathfrak{X}(t_{j}, j+1))&\leq U(\mathfrak{X}(t_{j}, j))+\alpha_{3}(|e_{\f}|)+\alpha_{4}(|e_{\rf}|),
\end{align}
where $\varpi\in(0, \min\{\mu, \theta_{1}, \theta_{2}\})$, $\zeta_{k}(v):=\max\{\bar{\zeta}_{k}(v), \bar{\sigma}_{k}(v)\}$ with $k\in\{1, 2, 3\}$, $\alpha_{3}(v):=\max\{\bar{\alpha}_{3}(v), \bar{\alpha}_{5}(v)\}$, and $\alpha_{4}(v):=\max\{\bar{\alpha}_{4}(v), \bar{\alpha}_{6}(v)\}$. Integrating \eqref{eqn-B15}-\eqref{eqn-B16} from $(0, 0)$ to $(t, j)$ in the hybrid time domain, one has
\begin{align}
\label{eqn-B17}
U(\mathfrak{X}(t, j))&\leq e^{-\varpi t}U(\mathfrak{X}(0, 0))+\frac{1}{\varpi(1-e^{-\varepsilon\varpi})}[\zeta_{1}(\|e_{\f}\|_{(t, j)})  \nonumber\\
&\quad +\zeta_{2}(\|e_{\rf}\|_{(t, j)})+\varpi\alpha_{3}(\|e_{\f}\|_{(t, j)}) \nonumber\\
&\quad +\varpi\alpha_{4}(\|e_{\rf}\|_{(t, j)})+\zeta_{3}(\|w\|_{(t, j)})],
\end{align}
where $\varepsilon:=\min_{i\in\mathcal{N}}\{\varepsilon_{i}\}$ and $\varepsilon_{i}$ is given in Assumption \ref{asn-3}. From \eqref{eqn-B4} and \eqref{eqn-B17}, we have
\begin{align*}
&|(\eta(t, j), e_{\ax}(t, j))|\leq\alpha^{-1}_{1}(2e^{-\varpi t}\alpha_{2}(|\mathfrak{X}(0, 0)|))\\
&\quad +\alpha^{-1}_{1}(4\bm{\zeta}_{1}(\|e_{\f}\|_{(t, j)})+\bm{\zeta}_{2}(8\|e_{\f}\|_{(t, j)})+\bm{\zeta}_{3}(8\|w\|_{(t, j)})),
\end{align*}
where $\bm{\zeta}_{1}(v):=(1-e^{-\varepsilon\varpi})^{-1}(\varpi^{-1}\zeta_{1}(v)+\alpha_{3}(v))$, $\bm{\zeta}_{2}(v):=(1-e^{-\varepsilon\varpi})^{-1}(\varpi^{-1}\zeta_{2}(v)+\alpha_{4}(v))$ and $\bm{\zeta}_{3}(v):=(1-e^{-\varepsilon\varpi})^{-1}\varpi^{-1}\zeta_{3}(v)$. Thus, the system \eqref{eqn-18} is ISS from $(e_{\rf}, e_{\f}, w)$ to $(\eta, e_{\ax})$ with $\beta(v, t, j):=\alpha^{-1}_{1}(2e^{-\varpi(0.5t+0.5\varepsilon j)}\alpha_{2}(v))$, $\gamma_{1}(v):=\alpha^{-1}_{1}(4\bm{\zeta}_{1}(v))$, $\gamma_{2}(v):=\alpha^{-1}_{1}(8\bm{\zeta}_{2}(v))$, and $\gamma_{3}(v):=\alpha^{-1}_{1}(8\bm{\zeta}_{3}(v))$, where the definition of $\beta$ comes from the fact that $t\geq\varepsilon j$ (see \cite{Carnevale2007lyapunov}).

\subsection{Proof of Theorem \ref{thm-2}}

For all $\mathfrak{X}\in C\cup D\cup G(D)$, define the Lyapunov function $U(\mathfrak{X}):=V(x)+\mathcal{W}(\mathfrak{X})$ with $\mathcal{W}(\mathfrak{X}):=\max\{\gamma_{b}\phi_{b}(\tau)W^{2}(e, m, \kappa, b), (1-b)\rho V(x)\}$. Similar to the proof of Theorem \ref{thm-1}, there exist $\alpha_{1}, \alpha_{2}\in\mathcal{K}_{\infty}$ such that \eqref{eqn-B4} holds for all $\mathfrak{X}\in C\cup D\cup G(D)$.

Consider the evolution of $U(\mathfrak{X})$ on the flow. If $\gamma_{b}\phi_{b}(\tau)W^{2}(\kappa, e)\geq\rho V(x)$, then we have from \eqref{eqn-39}-\eqref{eqn-40} that
\begin{align*}
&\langle\nabla U(\mathfrak{X}), F(\mathfrak{X}, w)\rangle\leq-\mu V(x)-\theta W^{2}(e, m, \kappa, b)+\Pi(\mathfrak{X}),
\end{align*}
where
\begin{align*}
\Pi(\mathfrak{X})&:=\gamma^{2}_{b}W^{2}(e, m, \kappa, b)-H^{2}_{b}(x, e)+\zeta_{1b}(|e_{\f}|)+\zeta_{2b}(|e_{\rf}|) \nonumber\\
&\quad +\zeta_{3b}(|w|)+\gamma_{b}[-2L_{b}\phi_{b}(\tau)\nonumber\\
&\quad-\gamma_{b}((1+\varrho_{b})\phi^{2}_{b}(\tau)+1)]W^{2}(e, m, \kappa, b) \nonumber\\
&\quad+2\gamma_{b}\phi_{b}(\tau)W(e, m, \kappa, b)[L_{b}W(e, m, \kappa, b)+H_{b}(x, e) \nonumber\\
&\quad+\sigma_{1b}(|e_{\f}|)+\sigma_{2b}(|e_{\rf}|)+\sigma_{3b}(|w|)] \nonumber\\
&\leq-(H^{2}_{b}(x, e)-\gamma_{b}\phi_{b}(\tau)W(e, m, \kappa, b))^{2} \nonumber\\
&\quad+\zeta_{1b}(|e_{\f}|)+\zeta_{2b}(|e_{\rf}|)+\zeta_{3b}(|w|) \nonumber\\
&\quad+3\varrho^{-1}_{b}[\sigma^{2}_{1b}(|e_{\f}|)+\sigma^{2}_{2b}(|e_{\rf}|)+\sigma^{2}_{3b}(|w|)],
\end{align*}
where, ``$\leq$'' holds due to the fact that $2\mathfrak{a}\mathfrak{b}\leq\mathfrak{c}\mathfrak{a}^{2}+\mathfrak{b}^{2}/\mathfrak{c}$ for all $\mathfrak{a}, \mathfrak{b}\geq0, \mathfrak{c}>0$. Hence,
\begin{align}
\label{eqn-B20}
\langle\nabla U(\mathfrak{X}), F(\mathfrak{X}, w)\rangle&\leq-\varpi U(\mathfrak{X})+\bar{\sigma}_{1}(|e_{\f}|) \nonumber\\
&\quad +\bar{\sigma}_{2}(|e_{\rf}|)+\bar{\sigma}_{3}(|w|)],
\end{align}
where $\varpi\in(0, \min\{\mu, \theta\})$, $\bar{\sigma}_{k}(v):=\max_{b\in\{0, 1\}}\{\zeta_{kb}(v)+3\varrho^{-1}_{b}\sigma^{2}_{kb}(v)\}$ with $k\in\{1, 2, 3\}$. If $\gamma_{b}\phi_{b}(\tau)W^{2}(e, m, \kappa, b)<\rho V(x)$, then $b=0$ and for the flow equation,
\begin{align*}
&\langle\nabla U(\mathfrak{X}), F(\mathfrak{X}, w)\rangle=(1+\rho)\langle\nabla V(x), f(\delta, x, e, w)\rangle \nonumber \\
&\quad \leq(1+\rho)[\bar{\Pi}(\mathfrak{X})+\zeta_{10}(|e_{\f}|)+\zeta_{20}(|e_{\rf}|)+\zeta_{30}(|w|)],
\end{align*}
where
\begin{align*}
\bar{\Pi}(\mathfrak{X})&:=-\mu V(x)+(\gamma^{2}_{0}-\theta)W^{2}(e, m, \kappa, 0) \nonumber\\
&\quad-H^{2}_{0}(x, e)+\zeta_{10}(|e_{\f}|)+\zeta_{20}(|e_{\rf}|)+\zeta_{30}(|w|)  \nonumber\\
&\leq-\mu V(x)+(\gamma^{2}_{0}-\theta)W^{2}(e, m, \kappa, 0) \nonumber\\
&\quad+\zeta_{10}(|e_{\f}|)+\zeta_{20}(|e_{\rf}|)+\zeta_{30}(|w|).
\end{align*}
Since $\gamma_{0}\phi_{0}(\tau)W^{2}(e, m, \kappa, 0)<\rho V(x)$, we obtain $\gamma^{2}_{0}W^{2}(e, m, \kappa, 0)<\gamma_{0}\rho\phi^{-1}_{0}(\tau)V(x)<(\max\{1, \mu\gamma^{-1}_{0}\})^{-1}$ $\cdot\min\{1, \mu\gamma^{-1}_{0}\}\mu V(x)<\pi\mu V(x)$, where $\pi\in(0, 1)$. Hence,
\begin{align}
\label{eqn-B21}
\langle\nabla U(\mathfrak{X}), F(\mathfrak{X}, w)\rangle&\leq-\tilde{\varpi}U(\mathfrak{X})+\zeta_{10}(|e_{\f}|) \nonumber\\
&\quad +\zeta_{20}(|e_{\rf}|)+\zeta_{30}(|w|),
\end{align}
where $\tilde{\varpi}\in(0, \min\{(1-\pi)\mu, \theta\})$.

Next, consider the evolution of $U(\mathfrak{X})$ at the jumps. For the case that $b=0$ and the ETM is applied, we have
\begin{align}
\label{eqn-B22}
U(G_{1}(\mathfrak{X}))&\leq V(x)+\gamma_{1}\phi_{1}(0)[\lambda W(e, m, \kappa, 0)\nonumber \\
&\quad+\alpha_{3W}(|e_{\f}|)+\alpha_{4W}(|e_{\rf}|)]^{2}\nonumber\\
&\leq V(x)+\gamma_{1}\phi_{1}(0)[(1+\varrho_{1})\lambda^{2}W^{2}(e, m, \kappa, 0)\nonumber\\
&\quad+(1+2\varrho^{-1}_{1})(\alpha^{2}_{3W}(|e_{\f}|)+\alpha^{2}_{4W}(|e_{\rf}|))]\nonumber\\
&\leq V(x)+\gamma_{0}\phi_{0}(\tau)W^{2}(e, m, \kappa, 0) \nonumber\\
&\quad+\bar{\alpha}_{3W}(|e_{\f}|)+\bar{\alpha}_{4W}(|e_{\rf}|)\nonumber\\
&=U(\mathfrak{X})+\bar{\alpha}_{3W}(|e_{\f}|)+\bar{\alpha}_{4W}(|e_{\rf}|),
\end{align}
where, the second ``$\leq$'' holds due to the fact that $2\mathfrak{a}\mathfrak{b}\leq\mathfrak{c}\mathfrak{a}^{2}+\mathfrak{b}^{2}/\mathfrak{c}$ for all $\mathfrak{a}, \mathfrak{b}\geq0, \mathfrak{c}>0$, the third ``$\leq$'' holds from \eqref{eqn-44}, $\bar{\alpha}_{3W}(v):=\gamma_{1}\phi_{1}(0)(1+2\varrho^{-1}_{1})\alpha^{2}_{3W}(v)$, and $\bar{\alpha}_{4W}(v):=\gamma_{1}\phi_{1}(0)(1+2\varrho^{-1}_{1})\alpha^{2}_{4W}(v)$. For the case $b=1$,
\begin{align}
\label{eqn-B23}
U(G_{2}(\mathfrak{X}))&=(1+\rho)V(x)=U(\mathfrak{X}).
\end{align}
For the case that $b=1$, we have
\begin{align}
\label{eqn-B24}
U(G_{1}(\mathfrak{X}))&\leq V(x)+\gamma_{0}\phi_{0}(\tau)[W(e, m, \kappa, 1)+\alpha_{5W}(|e_{\f}|) \nonumber \\
&\quad +\alpha_{6W}(|e_{\rf}|)]^{2}\nonumber\\
&\leq V(x)+\gamma_{0}\phi_{0}(\tau)[(1+\varrho_{0})W^{2}(e, m, \kappa, 1)\nonumber\\
&\quad+(1+2\varrho^{-1}_{0})(\alpha^{2}_{5W}(|e_{\f}|)+\alpha^{2}_{6W}(|e_{\rf}|))]\nonumber\\
&\leq V(x)+\gamma_{1}\phi_{1}(\tau)W^{2}(e, m, \kappa, 1) \nonumber\\
&\quad+\bar{\alpha}_{5W}(|e_{\f}|)+\bar{\alpha}_{6W}(|e_{\rf}|)\nonumber\\
&=U(\mathfrak{X})+\bar{\alpha}_{5W}(|e_{\f}|)+\bar{\alpha}_{6W}(|e_{\rf}|),
\end{align}
where, $\bar{\alpha}_{5W}(v):=\gamma_{0}\bar{\lambda}^{-1}(1+2\varrho^{-1}_{0})\alpha^{2}_{5W}(v)$ and $\bar{\alpha}_{6W}(v):=\gamma_{0}\bar{\lambda}^{-1}(1+2\varrho^{-1}_{0})\alpha^{2}_{6W}(v)$.

The remaining is the similar to the proof of Theorem \ref{thm-1}, and thus the system \eqref{eqn-18} is ISS from $(e_{\rf}, e_{\f}, w)$ to $(\eta, e_{\ax})$.

\end{document}